\documentclass[aps,prx,citeautoscript,twocolumn,superscriptaddress,floatfix,letterpaper,noamssymb,noamsmath,10pt,longbibliography]{revtex4-2}
\usepackage{silence}
\usepackage{physics}
\usepackage{dcolumn}
\usepackage{bm}
\usepackage{verbatim}
\usepackage[utf8]{inputenc} %use these 3 if you are compiling with pdflatex
\usepackage[LGR,T1]{fontenc} %use these 3 if you are compiling with pdflatex
\usepackage[english]{babel} %use these 3 if you are compiling with pdflatex
\usepackage[protrusion=true,expansion=true]{microtype}
\usepackage{tabularx}
\usepackage{fancyhdr}
\usepackage{float}
\usepackage{graphicx}
\usepackage{placeins}
\usepackage[margin=1in]{geometry}
\usepackage[toc,page]{appendix}
\usepackage[separate-uncertainty=true,multi-part-units=single]{siunitx}
\usepackage[breaklinks]{hyperref}

\let\sub\textsubscript

\begin{document}
\preprint{Preprint}
\title{Observation of a Zero-Field Josephson Diode Effect in a Helimagnet Josephson Junction}

\author{Alexander Beach}
\affiliation{Department of Physics and Materials Research Laboratory, University of Illinois Urbana-Champaign, United States}
\author{Mostafa Tanhayi Ahari}
\affiliation{Department of Natural Sciences and Math, Transylvania University, United States}
\author{Younghyuk Kim}
\affiliation{Department of Mechanical Science \& Engineering, University of Illinois Urbana-Champaign, United States}
\author{Kannan Lu}
\noaffiliation{}
\author{Gregory MacDougall}
\affiliation{Department of Physics and Materials Research Laboratory, University of Illinois Urbana-Champaign, United States}
\author{Matthew Gilbert}
\affiliation{Department of Electrical \& Computer Engineering, University of Illinois Urbana-Champaign, United States}
\author{Nadya Mason}
\affiliation{Pritzker School of Molecular Engineering, University of Chicago, United States}

\date{\today}

\begin{abstract}
We present Josephson junctions with the metallic chiral helimagnet Cr\sub{1/3}NbS\sub{2} as the weak link that exhibit magnetic diffraction patterns with asymmetry in both applied magnetic field and the critical (switching) current. The nonreciprocity between positive and negative critical currents (the Josephson diode effect) reaches efficiencies of $\abs{\eta}>20\%$, where $\eta \equiv (I_{c+}-|I_{c-}|)/(I_{c+}+|I_{c-}|)$, and the effect persists even at zero applied field. As a chiral helimagnet, Cr\sub{1/3}NbS\sub{2} lacks inversion symmetry and time reversal symmetry, providing a platform to explore the interplay of superconductivity with both symmetries broken. We propose that pinned Abrikosov vortices are a primary mechanism for the asymmetric magnetic-field response, while the nonzero spin chirality of Cr\sub{1/3}NbS\sub{2} gives rise to the diode effect. Simulations of magnetic diffraction patterns with vortices show offsets from zero field consistent with observations, while simulations of chiral spin structures with out-of-plane canting show a diode effect.
\end{abstract}

\maketitle

\clearpage
\newpage
\thispagestyle{fancy}

\section{Introduction}
The majority of studies of proximity superconductivity focus on systems with strong spin-orbit coupling, conventional ferromagnetism, and external magnetic fields. However, proximity superconductivity in systems with complex magnetic textures that host rich physics, such as antiferromagnets and altermagnets, remains relatively unexplored. In this manuscript we present transport measurements of Josephson junctions fabricated with the chiral helimagnet Cr\sub{1/3}NbS\sub{2}, and we observe both a Josephson diode effect that can persist at zero applied field and strongly shifted and hysteretic magnetic diffraction patterns.

\begin{figure*}[tbp]
    \centering
    \includegraphics[width=2\columnwidth]{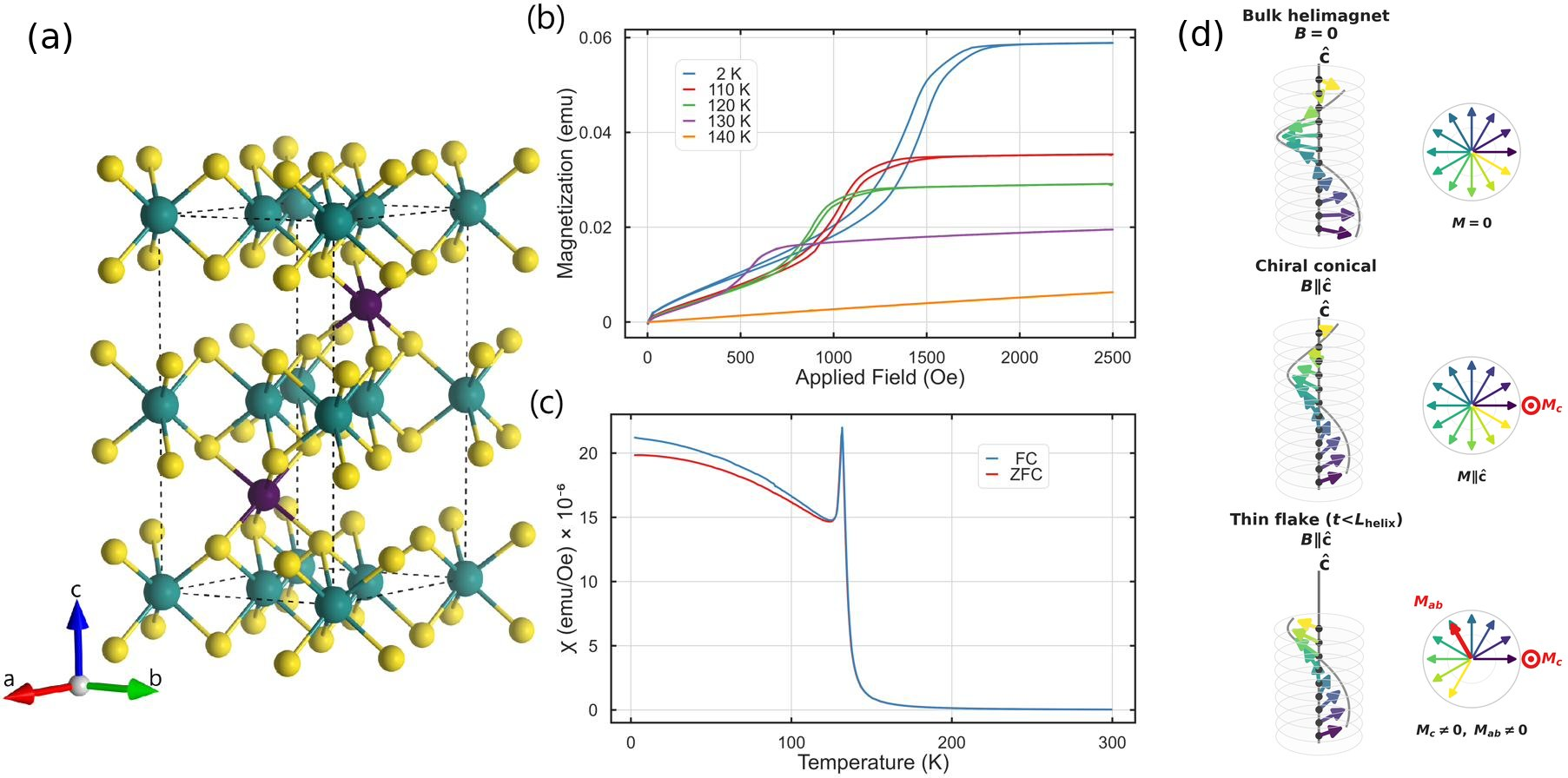}
    \caption{(a) The crystal structure of Cr\sub{1/3}NbS\sub{2}, Cr atoms (purple) are intercalated between the planes of NbS\sub{2} (blue, yellow). (b) Applied magnetic field vs magnetization data showing magnetic hysteresis in bulk Cr\sub{1/3}NbS\sub{2}. Above the magnetic transition temperature there is a purely paramagnetic response. The hysteresis in the measurements below the transition temperature is consistent with the presence of solitons. (c) Magnetic susceptibility data for field-cooled and zero-field-cooled bulk Cr\sub{1/3}NbS\sub{2}, with a clear transition at \SI{\sim 132}{\kelvin}. (d) Illustrations of the helimagnetic phases of Cr\sub{1/3}NbS\sub{2}. The bulk helimagnetic phase has near zero net magnetization, while the chiral conical phase has a net magnetization along the chiral axis due to an external field, $\mathbf{M}_{c}$. A flake thinner than the helical period will have its own net magnetization perpendicular to the chiral axis, $\mathbf{M}_{ab}$, and can cant along the chiral axis from an external field.}
    \label{MagneticCharacterization}
\end{figure*}

Compared to ferromagnets and antiferromagnets, Cr\sub{1/3}NbS\sub{2} has a complex, non-collinear spin texture. Above its magnetic phase transition at \SI{\sim 132}{\kelvin}, Cr\sub{1/3}NbS\sub{2} is paramagnetic, and below the transition the spin configuration becomes helimagnetic, with the helical axis pointing in the \textit{c}-direction of the crystal shown in \hyperref[MagneticCharacterization]{Figure 1a}, with a helical period of \SI{48}{\nano\meter}.\cite{MORIYA1982209} When an applied magnetic field is parallel to the helical axis, then the helimagnetic spin texture can become a chiral conical configuration, where the spins are canted toward the helical axis. The bulk helimagnetic and chiral conical configurations, together with the thin-flake regime relevant to our junctions in which the flake thickness is below one helical period, are illustrated in \hyperref[MagneticCharacterization]{Figure 1(d)}. However, a magnetic field applied perpendicular to the helical axis (i.e. within the ab-plane of the crystal) can give rise to one-dimensional solitons.\cite{CAO2020100080} Cr\sub{1/3}NbS\sub{2} is also a metallic transition metal dichalcogenide, made of NbS\sub{2} layers with intercalated Cr atoms in between the layers, and belongs to the non-centrosymmetric space group P6\sub{3}22.\cite{MORIYA1982209,Ghimire2013CrNbS2,Clements2017}

Simultaneous breaking of inversion symmetry (IS) by the non-centrosymmetric crystal structure and time-reversal symmetry (TRS) by magnetic order in Cr\sub{1/3}NbS\sub{2}, together with its distinct magnetic phase diagram, make it a good material for studying nonreciprocal effects. We observe that the Cr\sub{1/3}NbS\sub{2} Josephson junctions exhibit the Josephson diode effect, and that magnetic diffraction patterns from the junctions show an offset from zero magnetic field and non-monotonic amplitude decay. We propose that the helimagnetic spin texture of the Cr\sub{1/3}NbS\sub{2} creates the diode effect, based on simulations done with helical and chiral-conical spin textures. At the same time we propose that Abrikosov vortices in the leads of the junction are the cause of the offsets in magnetic field and unconventional magnetic diffraction patterns. Importantly, the simulated diode effect only occurs when the spins have some out-of-plane canting, i.e. when there is a chiral-conical configuration. To explain both the offset and the diode effect, we further suggest that there is an out-of-plane field from the Cr\sub{1/3}NbS\sub{2} that creates vortices in the leads of the junctions, possibly due to finite thickness of the Cr\sub{1/3}NbS\sub{2} flake compared to its helical period.

Two aspects make helimagnet-based Josephson junctions a particularly useful platform for nonreciprocal superconducting transport. First, the weak link provides intrinsic inversion-symmetry breaking and magnetic time-reversal-symmetry breaking, so a diode response can persist even when the externally applied field is returned to zero after magnetic training. Second, the same magnetic texture can generate local stray fields that nucleate and pin vortices, producing unconventional and hysteretic magnetic diffraction patterns. Observing both effects in a single material system helps separate phase-texture mechanisms from intrinsic current--phase-relation nonreciprocity and motivates helimagnets as a route to field-free superconducting rectification. In this work, we establish that 1) the magnitudes of the positive and negative critical currents differ and the diode response can persist at zero applied field after magnetic training, 2) the magnetic diffraction patterns are shifted from zero field, hysteretic, and show non-monotonic envelope evolution inconsistent with a uniform junction model, and 3) simulations show pinned Abrikosov vortices can reproduce offsets and distortions in $I_c(B)$ while chiral-conical spin textures yield a diode response, supporting a picture in which phase-texture effects and intrinsic nonreciprocity coexist.

The interplay between broken time reversal and spatial inversion symmetry, leading to nonreciprocal effects, was first deliberately observed in chiral molecules in an applied magnetic field, where it was found that the optical absorption (refractive index) of the molecules depended on their handedness.\cite{Rikken1997} The class of known nonreciprocal effects arising from broken time reversal and inversion symmetries is now quite large, with the effects arising from chiral broken inversion symmetry often called magnetochiral anisotropies.\cite{Barron20021984,Ideue2017,https://doi.org/10.1002/chir.23361,Legg2022,Rikken_Avarvari_2023,Qin2017} This has also come to include nonreciprocal charge transport in superconducting systems, prototypically thought of as the combination of a non-centrosymmetric superconductor with an applied magnetic field that breaks TRS.\cite{PhysRevB.109.054508,PhysRevB.103.245302,Bauriedl2022} In these systems it is possible for a supercurrent to flow in one direction, but in the opposite direction the current flow is dissipative. In other words, the magnitudes of the positive and negative critical currents are not the same, $I_{c+}\neq \abs{I_{c-}}$.\cite{nadeem2023superconductingdiodeeffect} This phenomenon is generally called the superconducting diode effect, and often the Josephson diode effect when the system is a Josephson junction. We quantify the nonreciprocity using the diode efficiency $\eta \equiv (I_{c+}-\abs{I_{c-}})/(I_{c+}+\abs{I_{c-}})$, where the sign of $\eta$ encodes the diode polarity (whether $I_{c+}$ or $|I_{c-}|$ is larger), and $\eta=0$ corresponds to a reciprocal device. 

Asymmetric critical current was observed as early as 1967, in wide, high-current Josephson junctions made of tin films.\cite{PhysRev.164.544} Schemes have also existed for producing a nonreciprocal supercurrent using arrays of Josephson junctions, long Josephson junctions, and intentionally designed fluxonic diodes.\cite{PhysRevE.61.2257,Falo2002,403209,PhysRevLett.87.077002} There are also now examples of asymmetric critical currents in systems without explicitly broken time-reversal symmetry, such as strained heterostructures and electrically polarized materials.\cite{doi:10.1126/sciadv.ado1502,PhysRevX.12.041013} More recently though, the superconducting diode effect has been observed in materials where the nonreciprocity arises from intrinsic material properties, rather than as a consequence of circuit design.

In superconductor/ferromagnet bilayers the diode effect has been attributed to asymmetric vortex dynamics arising from the stray field of the
ferromagnetic layer.\cite{Gutfreund2023-ht} The Josephson diode effect at zero field has been reported in non-centrosymmetric NbSe\sub{2}/Nb\sub{3}Br\sub{8}/NbSe\sub{2} Josephson junctions, in inversion-symmetric twisted tri-layer graphene and in multi-terminal graphene junctions.\cite{Wu2022,Lin-2022,PhysRevApplied.21.034011} In the former case the diode effect is attributed to polarization of the junction inducing asymmetric Josephson tunneling, while in the latter case the diode effect is suspected to come from a spin polarization or polarization of the valley occupations. Related field-free (and field-resilient) Josephson diodes have also been demonstrated in multiferroic Josephson junctions,\cite{Yang2025FieldResilient} and a Josephson diode based on a single conical magnet has been theoretically proposed.\cite{KamraFu2024Conical} The diode effect has also been observed in atomic Josephson junctions with a magnetic impurity, where it is attributed to dissipation caused by quasiparticle currents that have a particle-hole asymmetry.\cite{Trahms2023} To date there is no universal explanation for the superconducting diode effect in all systems, as the exact mechanism can be system specific.

\begin{figure*}[tb!]
    \centering
    \includegraphics[width=2\columnwidth]{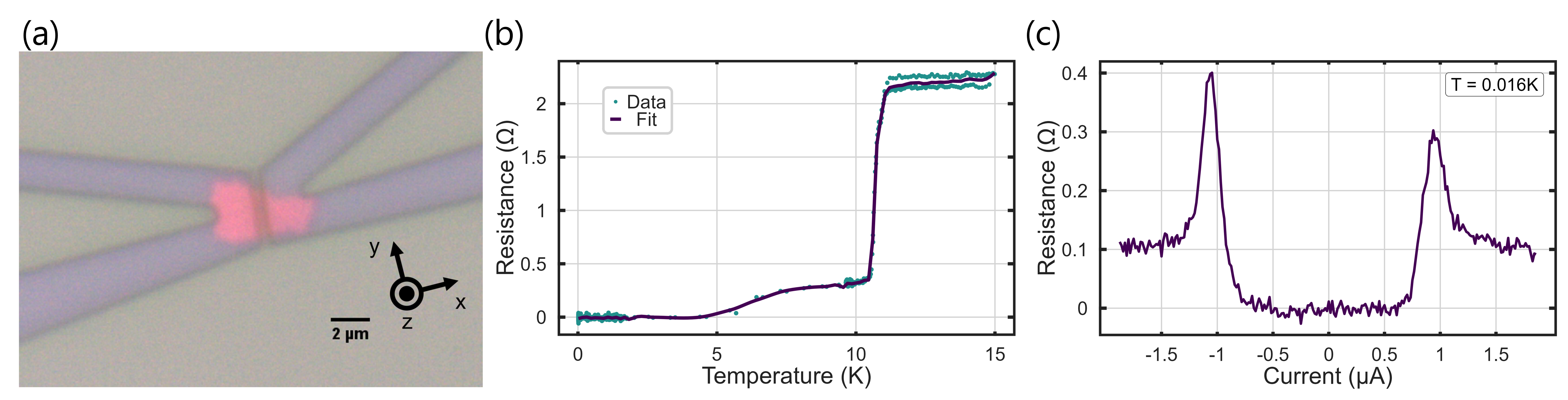}
    \caption{(a) An optical microscope image of a quasi-four-point Josephson junction pattern, after e-beam lithography development, but prior to NbTiN deposition, showing the Cr\sub{1/3}NbS\sub{2} weak-link region. (b) The resistance across the same junction as a function of temperature. The blue points are the raw data, and the red line is a fit. (c) The resistance of the junction as a function of applied current. This measurement was taken immediately after the initial cooldown to superconducting temperatures, with no applied magnetic field. There is a slight asymmetry in the positive and negative critical currents, as well as a difference in resistance peak amplitude at the critical currents.}
    \label{JunctionCharacteristics}
\end{figure*}

We now outline the structure of the paper. We first describe crystal growth, device fabrication, and the transport measurement configuration. We then
present the junction characteristics and magnetic diffraction patterns, emphasizing two key observations: (i) unconventional, hysteretic field offsets in the interference pattern and (ii) a pronounced Josephson diode effect that can persist to low applied field after magnetic training. Finally, we discuss the physical origins of these effects in terms of two complementary ingredients: phase-texture mechanisms (including vortex
contributions) that shift and distort the diffraction pattern, and an intrinsic nonreciprocal current-phase relationship associated with the chiral magnetic texture of the Cr\sub{1/3}NbS\sub{2} weak link.

\section{Experiment}

The Cr\sub{1/3}NbS\sub{2} used in these experiments was grown via chemical vapor transport with an iodine transport gas, which resulted in plate-like hexagonal crystals of about \SI{2}{\milli\meter} on a side. Representative crystals were characterized via single-crystal x-ray diffraction and Laue diffraction. Analysis of the diffraction data confirmed the P6\sub{3}22 space group with Cr$^{3+}$ atoms at the 2c site, consistent with the expected chiral structure, pictured in \hyperref[MagneticCharacterization]{Figure 1a}. Magnetization was measured using a SQUID magnetometer with applied field in the ab-plane (parallel to the large face of the crystal). \hyperref[MagneticCharacterization]{Figure 1b} shows that the magnetization has field hysteresis at temperatures below the magnetic transition temperature T\sub{N}, with the width of the hysteresis increasing with decreasing temperature, while above the transition temperature the behavior is paramagnetic. \hyperref[MagneticCharacterization]{Figure 1c} shows a sudden increase in magnetic susceptibility with decreasing temperature, with a sharp maximum of \SI{22E-6}{emu/Oe} at T\sub{N} = \SI{132}{\kelvin}. The magnetization hysteresis is caused by the formation of a soliton lattice, and the susceptibility peak is associated with the transition to a helical spin structure.\cite{Sirica2020,Clements2017}

\begin{figure*}[tbp]
    \centering
    \includegraphics[width=2\columnwidth]{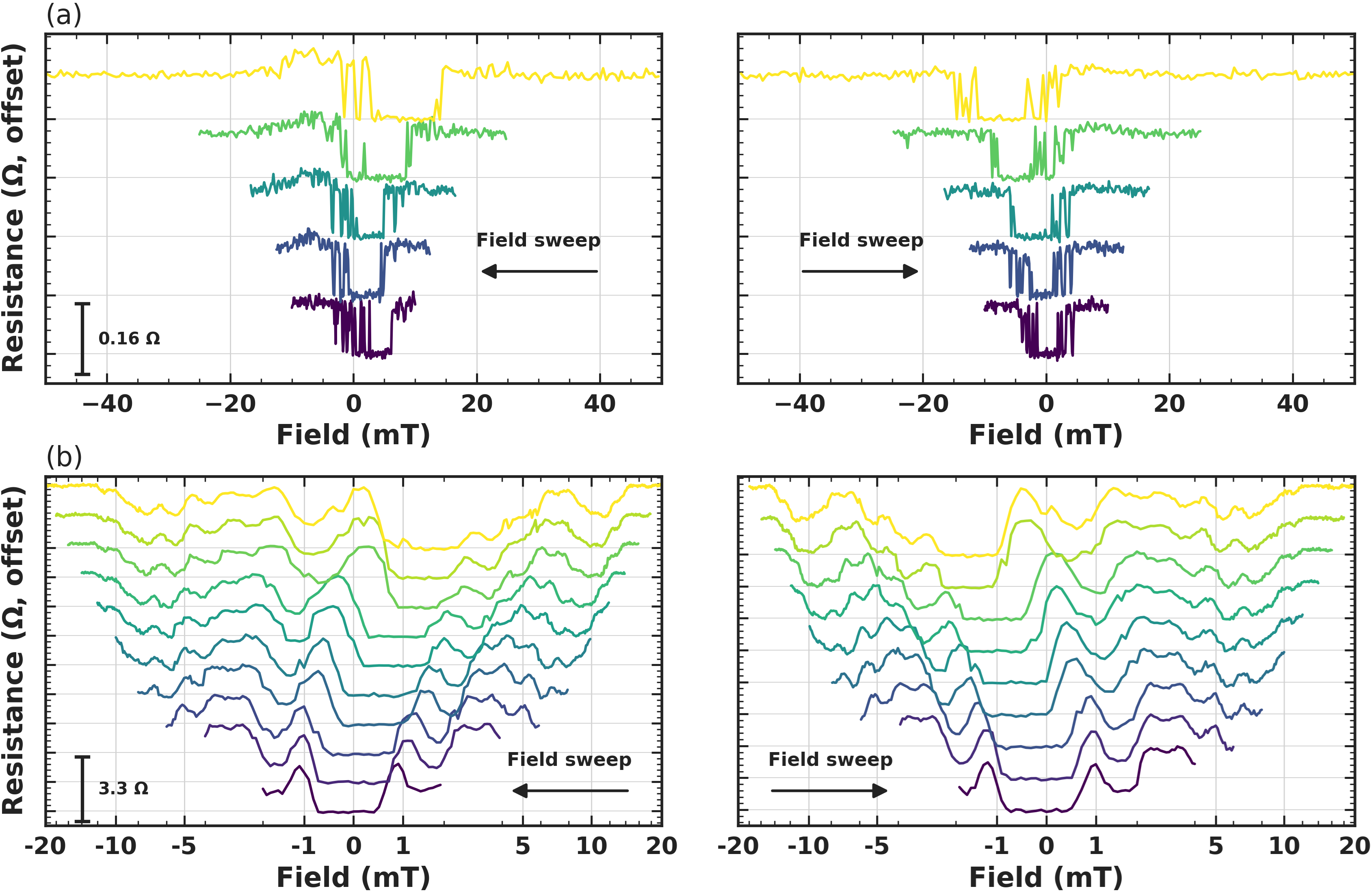}
    \caption{(a) The magnetoresistance of the junction as a magnetic field is applied perpendicular to the plane of the junction. The left plot shows magnetic field sweeps going from positive magnetic field to negative magnetic field, and the right plot shows sweeps from negative to positive. The lighter yellow curves begin at larger field magnitudes and are further offset from zero, while the darker purple curves begin at smaller magnetic field values and show less hysteresis. The curves are stacked with an arbitrary resistance offset for visual clarity. (b) Magnetoresistance as a function of out-of-plane magnetic field strength for a second junction. The hysteresis is more apparent when plotted on a logarithmic scale, and multiple dips in resistance are seen at larger field values. The reproducibility of the curves suggests that the hysteresis and local extrema are part of the larger magnetic diffraction pattern.}
    \label{JunctionCharacteristics2}
\end{figure*}

\begin{figure*}[tbp]
    \centering
    \includegraphics[width=2\columnwidth]{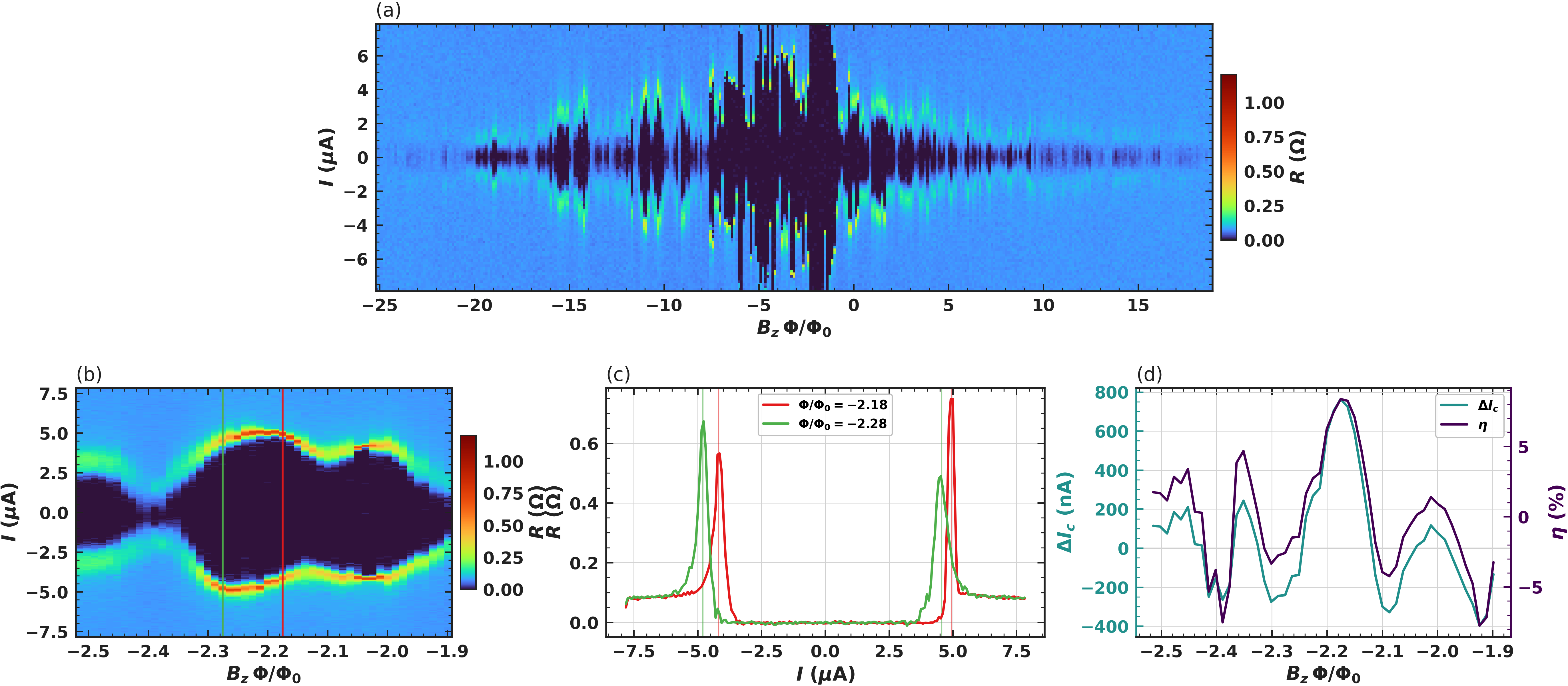}
    \caption{(a) Magnetic diffraction pattern in a Cr\sub{1/3}NbS\sub{2} junction for a field applied perpendicular to the plane of the junction. The field has been normalized to units of flux divided by flux quantum using the cross-sectional area of the junction. There are several notable features: there is a shift of the patterns largest peak from zero magnetic field, a large asymmetry with respect to field, a decay in peak amplitude that does not follow the standard $\sin(\pi \Phi/\Phi_0)/(\pi \Phi/\Phi_0)$ Fraunhofer behavior, and a diode effect. (b) A zoomed-in section of a magnetic diffraction measurement similar to a, making the diode effect clearer. Here the diode effect is visible as a tilting of the lobes in the pattern. (c) Line cuts of differential resistance at specific flux values from b. The red curve ($\Phi/\Phi_{0}=-2.18$) shows the diode effect with a larger magnitude positive critical current, while the green curve ($\Phi/\Phi_{0}=-2.28$) shows a larger magnitude negative critical current. Both of these are at negative field values within a single flux quantum of each other. (d) Diode-effect magnitude extracted from the data in b. The difference in critical currents $\Delta I_c(B) = I_{c+}(B) - |I_{c-}(B)|$ is plotted on the left vertical axis (teal), and the diode efficiency $\eta(B) = \Delta I_c(B)/(I_{c+}(B) + |I_{c-}(B)|)$ is plotted on the right vertical axis (purple). $\Delta I_c$ takes nonzero values across each diffraction lobe, and changes signs within and between lobes, with $\abs{\eta}$ reaching up to 8\% in this field window.}
    \label{MagneticDiffraction}
\end{figure*}

The junctions are fabricated from thin flakes of Cr\sub{1/3}NbS\sub{2} on Si substrates with \SI{500}{\nano\meter} oxides that were cleaned by oxygen reactive ion etching to enhance flake adherence. Successfully fabricating and coupling these types of materials to superconductors is generally difficult, and so yield was quite low. All successfully proximity-coupled flakes were under \SI{40}{\nano\meter} thick (below the
helimagnetic period of \SI{48}{\nano\meter}) and were identified via atomic force microscopy. We observed a diode effect across multiple junctions in this thickness regime; however, the available thickness range and number of devices are too small to determine a quantifiable dependence of $\eta$ on thickness. Electron beam lithography was used to pattern the junctions with quasi-four-point leads, shown in \hyperref[JunctionCharacteristics]{Figure 2a}, with at least \SI{60}{\nano\meter} of DC plasma sputtered NbTiN. The samples were cooled to \si{\milli\kelvin} temperatures in a dilution refrigerator with a 6:1:\SI{1}{\tesla} vector magnet. Electrical measurements were done with a combination of an SR830 lock-in amplifier, Keithley 2400 voltage source, Keithley 6221 current source, DC-DAQ, and custom electronics.

An optical microscope image of one of the junctions without NbTiN is shown in \hyperref[JunctionCharacteristics]{Figure 2a}, where the dimensions of this junction are roughly \SI{0.4}{\micro\meter} $\times$ \SI{2.6}{\micro\meter} $\times$ \SI{20}{\nano\meter}. The $\hat{x}$- and $\hat{y}$-directions are in the \textit{ab}-plane of the crystal, which is also the plane of the junction. $\hat{x}$ is defined as parallel to the current running through the junction, the $\hat{y}$-direction is defined as perpendicular to that current, and the $\hat{z}$-direction is perpendicular to the plane of the junction and is equivalent to the \textit{c}-direction of the Cr\sub{1/3}NbS\sub{2}. \hyperref[JunctionCharacteristics]{Figure 2b} shows an initial superconducting transition for the NbTiN leads at around \SI{10.7}{\kelvin}, with a zero resistance state for the junction below \SI{4.6}{\kelvin}. At temperatures above the NbTiN transition, the device has a finite resistance, consistent with a conducting (rather than insulating) weak link. \hyperref[JunctionCharacteristics]{Fig.~2c} shows the switching (critical) current for the same junction at \SI{16}{\milli\kelvin} with no applied magnetic field. The positive critical current is \SI{0.94}{\micro\ampere}, while the negative critical current is \SI{-1.05}{\micro\ampere}. These measurements indicate that the junction is well proximitized and there is a zero-field Josephson diode effect: the magnitude of the negative-bias critical current is about 11.7\% larger than the positive-bias critical current, with $\eta \approx -5.5\%$.

\begin{figure*}[tbp]
    \centering
    \includegraphics[width=2\columnwidth]{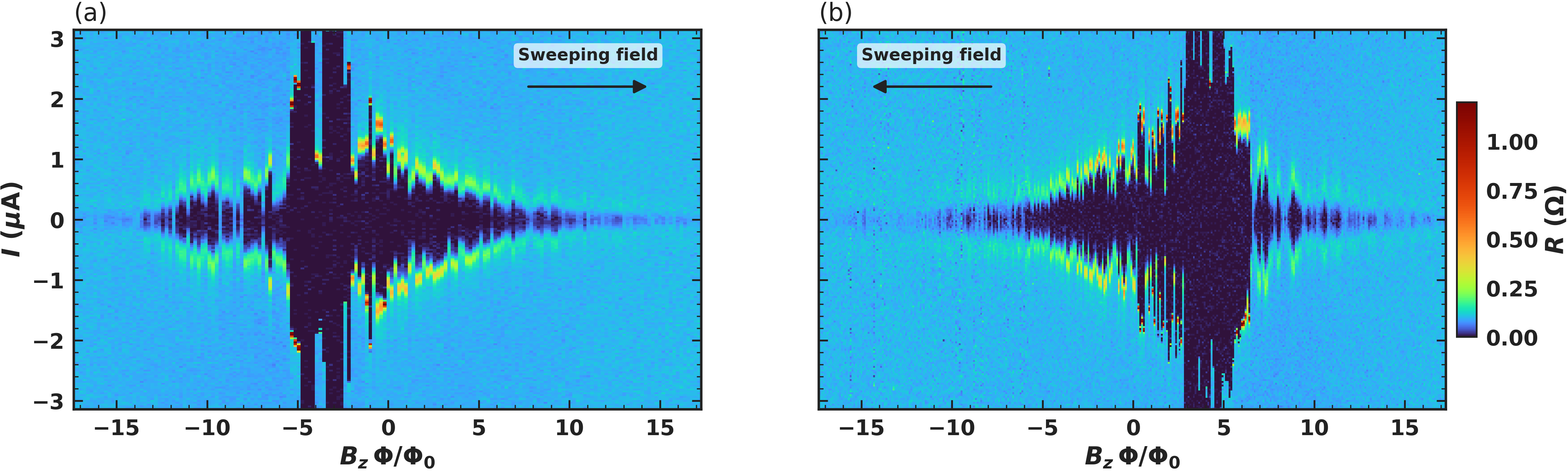}
    \caption{Magnetic diffraction patterns with a magnetic field sweeping from negative to positive, and positive to negative. The patterns are not perfect mirror images of each other, but do have the same overall shape and features. The central node shift away from zero field is not consistent with ferromagnetic hysteresis, where a positive (negative) coercive field is necessary to return from negative (positive) remnant magnetization to zero magnetization.\cite{10.1063/5.0195229}}
    \label{FullFraun}
\end{figure*}

\section{Results}
The magnetotransport measurements of the Cr\sub{1/3}NbS\sub{2} junctions show several phenomena, including anomalous magnetic hysteresis, magnetic field asymmetry, magnetic field aperiodicity, and a diode effect. All of these phenomena result in magnetic diffraction pattern measurements that are significantly distorted from an ideal Fraunhofer pattern.

\hyperref[JunctionCharacteristics2]{Figure 3} shows resistance measurements in the junction with out-of-plane magnetic field sweeps of varying magnitude and direction. In the left panels of \hyperref[JunctionCharacteristics2]{Figures 3a and 3b} the magnetic field is swept from positive to negative and vice versa in the right panels, with the yellow curves starting at \SI{\pm0.05}{\tesla}. Darker curves start at smaller field magnitudes and are shifted less. There is a clear hysteresis in the critical field gap, and no sweep with a superconducting region centered at zero field. This is more apparent in \hyperref[JunctionCharacteristics2]{Figure 3b} where the magnetic field scale is logarithmic. The size of the critical field gap also increases with the strength of the starting field. The observed hysteretic behavior is not consistent with standard ferromagnetic hysteresis. In a weak ferromagnet a large positive field strongly magnetizes the material and simultaneously weakens superconductivity. As the magnetic field is brought towards zero, the remnant magnetization should continue to weaken superconductivity, and to return to an un-magnetized state a small negative field must be applied to unpin the positive remnant magnetization, and so the proximitized superconductivity should be strongest (the critical current should be largest) at a small negative field. We report an opposite response, where a large positive (negative) magnetizing field does not need to be brought back to a negative (positive) field to recover a superconducting state; in fact a small positive (negative) field accomplishes this. Ferromagnetic Josephson junctions typically show either this expected behavior or negligible hysteresis.\cite{PhysRevB.80.220502,Ai2021,Qiu2023}

By applying a magnetic field and DC bias current together, the magnetic hysteresis from \hyperref[JunctionCharacteristics2]{Fig.~3} presents as a magnetic diffraction pattern that is offset from zero field. \hyperref[MagneticDiffraction]{Fig.~4a} shows that nonstandard features appear when a magnetic field is applied in the $\hat{z}$ direction, i.e., perpendicular to the applied current through the junction and parallel to the chiral axis of Cr\sub{1/3}NbS\sub{2}. \hyperref[MagneticDiffraction]{Figure 4(b)} is a zoomed-in section of a diffraction measurement similar to \hyperref[MagneticDiffraction]{(a)}, around a few diffraction lobes. It is visible in \hyperref[MagneticDiffraction]{(b)} that the diffraction lobes appear tilted, leading to a diode effect that changes sign in a single lobe. \hyperref[MagneticDiffraction]{Figure 4(c)} is differential resistance data at two selected magnetic field values, marked by red and green vertical lines in (b). From that data we can extract the positive-bias and negative-bias critical currents, $I_{c+}(B)$ and $I_{c-}(B)$, and quantify the nonreciprocity by the difference $\Delta I_c(B)=I_{c+}(B)-|I_{c-}(B)|$ (equivalently, by the diode efficiency $\eta(B)=\Delta I_c(B)/(I_{c+}(B)+|I_{c-}(B)|)$). The diode polarity is set by the sign of $\Delta I_c(B)$, and depending on field, either $I_{c+}$ or $|I_{c-}|$ can be larger. Importantly, the polarity does not simply track the sign of $B$. Within a given diffraction lobe, $\Delta I_c(B)$ can change sign, and similar behavior is observed for lobes at both positive and negative fields. \hyperref[MagneticDiffraction]{Figure 4(d)} shows the $\Delta I_c(B)$ and $\eta(B)$ extracted directly from \hyperref[MagneticDiffraction]{Figure 4(b)}, with the nonreciprocity oscillating above and below zero. While not shown, extracting $\eta$ from the data in \hyperref[MagneticDiffraction]{Figure 4(a)} yields values in excess of 20\%.

Joule heating can mimic diode-like behavior by producing an apparent asymmetry that depends on the current sweep direction. We rule this out for two reasons. First, the extracted $I_{c\pm}(B)$ are independent of whether the bias is swept from negative to positive or vice versa. Second, the observed reversals of diode polarity as a function of field occur without changing the sweep direction, consistent with a field-dependent modification of the Josephson response rather than a sweep-history-dependent thermal artifact.

The magnetic diffraction patterns in \hyperref[FullFraun]{Figure 5} show multiple features that deviate from the standard $\sin(\pi \Phi/\Phi_0)/(\pi \Phi/\Phi_0)$ Fraunhofer behavior seen in rectangular SNS Josephson junctions. The node spacing of the patterns is aperiodic, rather than being spaced at integer flux quanta values. Note that the flux quanta values in the axes of \hyperref[FullFraun]{Figure 5} are scaled from $\Phi=BA$, where $B$ is the value of the applied magnetic field, and $A$ is the actual cross-sectional area of the junction perpendicular to the field, and does not account for any flux focusing effects. In addition to aperiodicity the antinodal critical currents for larger applied fields do not monotonically decay, which can be seen clearly in \hyperref[FullFraun]{Figure 5}. Patterns with anomalous features have also been reported in other devices such as semiconductor Josephson junctions, topological insulator junctions, and ferromagnet junctions.\cite{PhysRevB.95.035307,Mayer2020,Assouline2019,Ai2021} It is known that asymmetry in magnetic diffraction patterns can be caused by slips in the superconducting phase difference across a junction and anisotropic magnetic susceptibility.\cite{Beach2021} Such a phase slip can be caused by external magnetic fields or geometric defects in the junction. For junctions made from mechanically exfoliated materials if the junction is not rectangular or the height of the flake in the junction varies sharply, then the Rashba coefficient, which characterizes the strength of spin-orbit coupling, could vary across the junction. However, the junctions used in this experiment show no geometric defects in atomic force microscopy scans. Because we see these unconventional Fraunhofer features in multiple devices, with a coexisting Josephson diode effect, we lean towards more intrinsic material-specific mechanisms to explain these results. The next section will discuss the origin of the anomalous features in these helimagnet junctions, and reconcile them with the presence of the Josephson diode effect.

\section{Origins of the Diode Effect and Magnetic Anomalies}

\begin{figure}[bp]
  \includegraphics[width=\columnwidth]{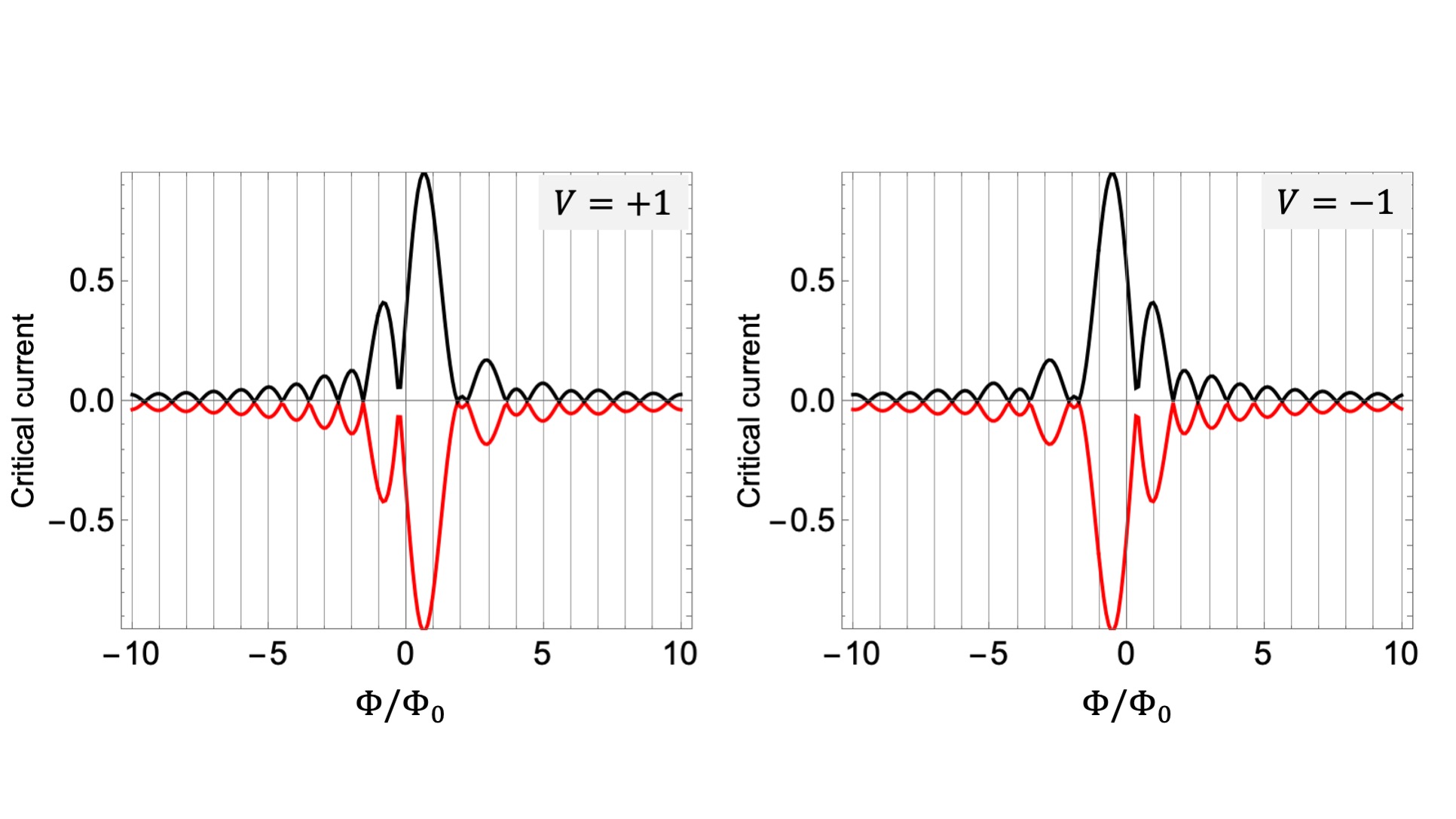}
   \caption{Magnetic diffraction patterns for vortex (left) and antivortex (right) at $(x_v,y_v)=(0.1,0)$. Here, we set $L_y=\Phi_0=I_c=1$ and ignore the vortex stray fields, $\Phi_v=0$.} 
   \label{vortexF}
\end{figure}

We propose two complementary explanations for the different features present in the data. First, the magnetic diffraction pattern $I_c(B_z)$ is primarily determined by the spatial phase profile across the junction, $\phi(y)$, and any mechanism that introduces an additional phase gradient or phase texture (e.g., trapped Abrikosov vortices, screening currents, or self-field/inductive effects in the leads) can shift and distort the interference pattern and produce history-dependent offsets in field. Such phase reconfiguration effects have been shown to generate competing Josephson-vortex states and diode-like nonreciprocity in thin-film junctions.\cite{Chen2024HiddenStates} Second, the intrinsic nonreciprocity between positive- and negative-bias switching currents in a Josephson diode effect is most naturally expressed as a nonreciprocal current-phase relation (CPR).\cite{PhysRevX.12.041013} In our devices these two phenomena can coexist, with phase-texture effects shaping the field dependence and lobe-to-lobe behavior of the measured nonreciprocity, while the chiral magnetic texture of Cr\sub{1/3}NbS\sub{2} provides a route to an intrinsically nonreciprocal CPR (and in particular with spins in a conical state with a finite out-of-plane moment). Accordingly, in this section we first discuss phase-texture mechanisms that can account for the observed offsets and asymmetries in the magnetic diffraction patterns, and then discuss how the chiral spin texture of Cr\sub{1/3}NbS\sub{2} can generate a diode effect via a nonreciprocal CPR.

\begin{figure}[bp]
  \includegraphics[width=\columnwidth]{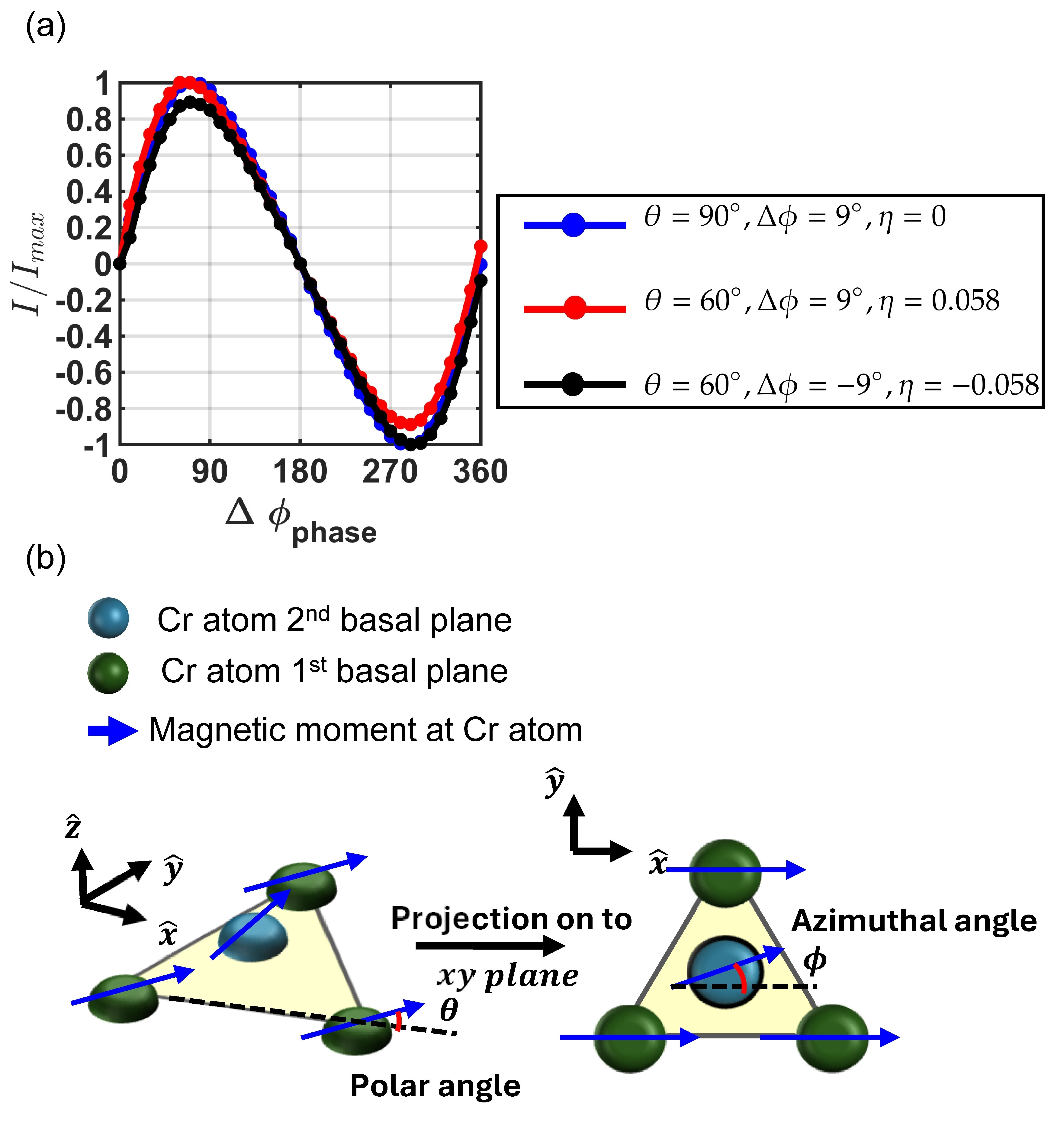}
\caption{(a) Normalized current--phase relationships (CPRs) for different spin textures calculated from the model described in the \hyperref[AppendixA]{Appendix}. The Josephson diode effect is quantified by the (signed) diode efficiency $\eta \equiv (I_{c+}-|I_{c-}|)/(I_{c+}+|I_{c-}|)$, where $I_{c+}$ and $|I_{c-}|$ are the magnitudes of the positive and negative critical currents extracted from each CPR. The helical texture with spins confined to the $ab$ plane ($\theta = 90^\circ$ and $\Delta\phi = 9^\circ$, where $\Delta\phi$ is the inter-layer azimuthal rotation increment of the magnetic texture) yields a symmetric CPR and thus $\eta = 0$ (blue). A conical texture ($\theta = 60^\circ$ and $\Delta\phi = 9^\circ$) yields a finite diode effect with $\eta \approx 5.8\%$ (red), while reversing the chirality ($\theta = 60^\circ$ and $\Delta\phi = -9^\circ$) reverses the diode polarity, $\eta \approx -5.8\%$ (black). Currents are normalized to $I_{\text{max}}$. (b) Diagram depicting the relative directions of the magnetic moments on Cr atoms in different planes. The diode effect only appears when there is an out-of-plane component to the moments, i.e. along the crystallographic \textit{c}-direction of Cr\sub{1/3}NbS\sub{2}.}
\label{cpr}
\end{figure}

The presence of an Abrikosov vortex can induce a phase difference in the junctions, through vortex currents and vortex fields, each dependent on the vortex core's proximity to the junction and the junction's geometry~\cite{Krasnov,Clem}. For thick superconductors vortex currents circulating around the core decay with a characteristic length $\lambda$ known as the London penetration length. In contrast, for a thin superconducting film $d\ll \lambda$, where $d$ is the thickness of the superconductor in the $\hat{z}$-direction, the characteristic length is given by the Pearl length $\lambda_P=\lambda^2/d$. When the vortex distance from the junction is comparable to the Pearl length, these currents cause a phase difference
\begin{align}\label{vEq}
\phi_v(y) = -V \arctan \frac{y - y_v}{|x_v|},
\end{align}
across the junction. Here, $(x_v,y_v)$ denotes the position of the vortex in the lead, and $V=+1$ ($V=-1$) corresponds to a vortex (antivortex). In contrast, when the vortex distance from the junction is larger than the Pearl length, the vortex current effects on the junction may be negligible. However, the vortex may have a long-range influence on the junction due to its stray magnetic fields~\cite{PhysRevB.100.174511}. These fields induce an effective magnetic flux within the junction, opposing the vortex's polarity. Due to flux-focusing effects in the junction region, this field can be significant, leading to a shift in the Fraunhofer pattern. Although the long-range vortex field and the short-range vortex circulating current are distinct effects, they can lead to a similar phase difference across the junction~\cite{PhysRevB.100.174511}. In the presence of a vortex, we obtain the magnetic interference pattern, i.e., the Fraunhofer pattern, as

\begin{align}
    I_{c+}(\Phi) &= \max_{\phi}\!\left[I_{\text{tot}}(\phi,\Phi)\right],\\
    I_{c-}(\Phi) &= \min_{\phi}\!\left[I_{\text{tot}}(\phi,\Phi)\right].
\end{align}

More generally, the total current can be written in terms of an arbitrary current-phase relationship (CPR) $I_J(\vartheta)$ as:

\begin{align}
I_{\text{tot}}(\phi,\Phi) = \frac{1}{L_y}\int\limits_{-L_y/2}^{L_y/2}
I_J\!\left(\vartheta(y)\right)\, \dd y.
\label{eq:Itot_generalCPR}
\end{align}

For a conventional short junction $I_J(\vartheta)=I_c\sin\vartheta$. In the vortex simulations below we use this sinusoidal CPR in order to isolate the effect of the vortex-induced phase texture on the magnetic diffraction pattern. A non-sinusoidal (and potentially nonreciprocal) CPR can be incorporated by using the corresponding $I_J(\vartheta)$ in Eq.~\eqref{eq:Itot_generalCPR}, and for completeness, Eq.~\eqref{app:combinedCPR} writes the combined formalism explicitly. The local phase $\vartheta(y)$ is defined as,

\begin{align}
    \vartheta(y) = \phi + 2\pi \frac{\Phi}{\Phi_0}\frac{y}{L_y} + \phi_v(y).
\end{align}

Here, $L_y$ is the lateral dimension of the junction perpendicular to the current direction.

\hyperref[vortexF]{Figure 6} shows the simulated Fraunhofer patterns for a (anti)vortex $V=(-)+1$ at $(x_v,y_v)=(0.1,0)$, with $L_y=\Phi_0=I_c=1$. Note that
although $I_{c+}(\Phi)=|I_{c-}(\Phi)|$, the simulated Fraunhofer patterns show similar features to the measured Fraunhofer patterns, including: finite offsets from zero magnetic field, non-monotonic decay in the amplitude of the higher field lobes, and asymmetry about the central node. The phase difference given in Eq.~\eqref{vEq} results in central peak shifts for the vortex and anti-vortex that are compatible with the experimental Fraunhofer patterns in \hyperref[FullFraun]{Figure 5} (sweeping the field from a large positive field to zero and a large negative field to zero). Notably, a ferromagnetic domain in Cr\sub{1/3}NbS\sub{2} could also shift the Fraunhofer pattern by contributing an additional internal field offset. However, in our observations the critical current maximum occurs before the magnetic field sweeps to zero, which is difficult to reconcile with a simple internal-field shift and is more naturally explained by a history-dependent phase texture such as a trapped vortex configuration.

A key question is whether vortices are spontaneously created or explicitly nucleated by external magnetic fields. In our measurements the field-axis
offsets and asymmetries depend strongly on sweep history, which is naturally explained if vortices are nucleated at large $|B|$ and can remain trapped in the NbTiN leads as the field is reduced, thereby imposing a history-dependent phase texture across the junction. We suggest that the magnetization of Cr\sub{1/3}NbS\sub{2} is the source of such vortices in our devices, through a combination of a finite out-of-plane magnetic moment when the material is in a chiral conical spin configuration and stray fields from the edges of the exfoliated flake (see \hyperref[MagneticCharacterization]{Figure 1(d)}). Previous simulations have shown that in a superconductor/helimagnet bilayer system, where the chiral axis is in the plane of the superconductor, the helical magnetic field can induce vortices in the superconductor.\cite{Fukui_2018} It has also been reported that vortex pinning at the edges of a superconducting film can produce a diode effect in superconducting thin films.\cite{PhysRevLett.131.027001} In our geometry, such vortices could be promoted by fringing fields near the edges of the Cr\sub{1/3}NbS\sub{2} flake (or by small applied fields that fringe around non-ideal edges of the superconducting leads). Below the magnetic transition temperature, the Cr moments are approximately ferromagnetically aligned within each layer, while the inter-layer twist produces the helimagnetic texture; the edge of an exfoliated flake can therefore host geometry-dependent fringing fields with a finite out-of-plane component. Another explanation for the anomalous hysteresis and asymmetry in the magnetic diffraction patterns is the presence of chiral domain walls in the crystal; the dynamics of chiral domain walls under an applied magnetic field can produce effects similar to vortex-induced phase textures.\cite{doi:10.1126/science.1133239, Bouhon_2010,Tsuruta2016PRB}

While pinned Abrikosov vortices can explain the magnetic field shifts, aperiodicity, and non-monotonic decay in the magnetic diffraction patterns, they do not by themselves account for the Josephson diode effect. A Josephson diode effect requires broken inversion symmetry and broken time-reversal symmetry and can be viewed as arising from an asymmetric current--phase relation (CPR) with $I_{c+}\neq |I_{c-}|$.\cite{doi:10.1126/sciadv.abo0309,Pal2022, PhysRevX.12.041013} Because these symmetry breakings are often tuned by external electric or magnetic fields, the diode effect frequently appears only under applied fields; however, Cr\sub{1/3}NbS\sub{2} provides intrinsic symmetry breaking through its chiral crystal structure and magnetic order. The P6\sub{3}22 space group of Cr\sub{1/3}NbS\sub{2} is non-centrosymmetric.

In addition, the chiral spin structure induces effective spin-orbit coupling and thus breaks inversion symmetry. The chiral magnetic structure of
Cr\sub{1/3}NbS\sub{2} breaks TRS, and for flakes whose thickness is comparable to, or smaller than, the chiral period of \SI{48}{\nano\meter}, the cancellation over one full period can be incomplete, and stray fields near edges can produce a nonzero local stray field in the vicinity of the junction. Such a local out-of-plane field can also provide a Zeeman interaction, which in combination with spin-orbit coupling can contribute to an asymmetric CPR and thus to a diode response. This does not require ferromagnetic domain formation: finite-size effects and a conical canting component can generate local fringing fields even when the underlying texture remains non-collinear. In this sense, a nonzero local out-of-plane field can coexist with the helimagnetic ordering, and the sweep-history effects can arise from changes in trapped vortex configurations rather than conventional domain switching. Moreover, when the spins adopt a chiral conical configuration, there is a finite magnetization component along the crystallographic \textit{c}-axis, which provides an intrinsic source of time-reversal-symmetry breaking relevant for a Josephson diode effect. Although the chiral conical structure is not the zero-field ground state for bulk Cr\sub{1/3}NbS\sub{2}, it can be induced by a magnetic field. Additionally, strain is also known to stabilize new magnetic phases in Cr\sub{1/3}NbS\sub{2}, which could be relevant in proximitized devices due to metallization and thermal stresses upon cooldown.\cite{paterson2020tensile} Non-centrosymmetric superconductors can also host spin-singlet and spin-triplet pairing in the presence of magnetic ordering, and there is evidence for spin-triplet pairing in superconductor/Cr\sub{1/3}NbS\sub{2} heterostructures, which can also lead to a diode effect.\cite{PhysRevLett.93.027003,annurev:NoncentrosymmetricSuperconductors, PhysRevResearch.6.L012046,PhysRevLett.132.216001}

We focus on explaining the diode effect via the chiral conical phase of Cr\sub{1/3}NbS\sub{2}. To understand how the spin structure could induce the diode effect, we study the current-phase relationship in a Josephson junction with a chiral spin structure using a tight-binding and non-equilibrium Green's function formulation. The crystal lattice of the simulated junction is designed to preserve IS, so that any nonreciprocity in the CPR is attributable to a chiral spin texture. The spin orientation is represented by $\hat{\mathbf{n}}_{j} = \left(\sin(\theta)\cos(j\Delta\phi), \sin(\theta)\sin(j\Delta\phi), \cos(\theta)\right)$, where $\theta$ is the polar angle between the \textit{z}-axis and the spin, $\Delta\phi$ is the inter-layer azimuthal rotation increment, $j$ is the number of layers along the $\hat{z}$-direction, and so the azimuthal angle of the spin in layer $j$ is $j\Delta\phi$. We demonstrate in \hyperref[cpr]{Figure 7} that the junction exhibits the diode effect when $\theta = \ang{60}$, where the spin structure is in a chiral conical configuration. Conversely, the junction does not maintain the diode effect when $\theta = \ang{90}$, where the spin structure is in a chiral helical configuration, with no out-of-plane canting. Detailed results and the methodology for the supercurrent calculation are provided in the \hyperref[AppendixA]{Appendix}. 

To summarize, the Cr\sub{1/3}NbS\sub{2} Josephson junctions studied in this experiment show both a Josephson diode effect and magnetic diffraction patterns with pronounced asymmetry and offsets from zero magnetic field. We suggest that the primary cause of the Fraunhofer anomalies is the presence of one or more Abrikosov vortices. The diode polarity is not simply set by the sign of the applied out-of-plane field $B_z$. Within a single interference lobe we observe polarity reversals, with regimes where $I_{c+} > |I_{c-}|$ and nearby regimes where $I_{c+} < |I_{c-}|$. This differs from field-controlled Josephson diodes, in which reversing the field reverses the diode polarity, and it suggests a relation between the broken inversion symmetry (IS) and broken time-reversal symmetry (TRS) caused by the helimagnetism of the Cr\sub{1/3}NbS\sub{2}. In the typical Rashba case, IS-breaking defines a polar axis and TRS is broken along an orthogonal direction, creating a diode along their cross product. In contrast, in Cr\sub{1/3}NbS\sub{2} both broken symmetries are structured along $\hat{z}$. TRS is broken by the out-of-plane net moment, and IS is broken by the handedness of the chiral winding rather than by a polar direction. The diode effect is still seen along the current direction $\hat{x}$, because the polar axis follows from the chiral combination $\hat{z} \times \nabla_z \hat{n}.$\cite{TokuraNagaosa2018} The required out-of-plane moment to break TRS along $\hat{z}$ can arise when the helimagnetic spins adopt a chiral conical configuration, and simulations of a Josephson junction with such a structure show a diode effect. Turning to the Fraunhofer anomalies, simulations of a Josephson junction with a phase difference from a single vortex or anti-vortex produce patterns that are asymmetric about their central lobe and offset from zero magnetic field. Both the chiral conical spin structure and the presence of vortices can be caused by a magnetic field perpendicular to the plane of the junction. While some preliminary evidence suggests that the vortices are generated by the chiral field of the helimagnet, further studies are needed to disentangle intrinsic effects from potential external factors.

\section*{Acknowledgments}
This work was supported by NSF-BSF under DMR-2422090 and by the U.S. Department of Energy, Office of Science, National Quantum Information Science Research Centers. The experiment was carried out using some facilities of the Illinois MRSEC, supported by NSF DMR-1720633. X-ray diffraction measurements were taken in the George L. Clark X-ray Facility in the 3M Materials Chemistry Laboratory, at the University of Illinois Urbana-Champaign, while magnetization measurements were done in the University of Illinois Urbana-Champaign Materials Research Laboratory using a Quantum Design MPMS3 SQUID magnetometer. Laue diffraction measurements were also taken in the Materials Research Laboratory.

\section*{Data Availability}
All data supporting the findings in this study can be accessed via Reference \cite{AlexanderBeach2025}.

\onecolumngrid
\bibliography{references}

@article{Krasnov,
  title = {Detection of the Phase Shift from a Single Abrikosov Vortex},
  author = {Golod, T. and Rydh, A. and Krasnov, V. M.},
  journal = {Phys. Rev. Lett.},
  volume = {104},
  issue = {22},
  pages = {227003},
  numpages = {4},
  year = {2010},
  month = {Jun},
  publisher = {American Physical Society},
  doi = {10.1103/PhysRevLett.104.227003},
  url = {https://link.aps.org/doi/10.1103/PhysRevLett.104.227003}
}

@article{Clem,
  title = {Effect of nearby Pearl vortices upon the ${I}_{c}$ versus $B$ characteristics of planar Josephson junctions in thin and narrow superconducting strips},
  author = {Clem, John R.},
  journal = {Phys. Rev. B},
  volume = {84},
  issue = {13},
  pages = {134502},
  numpages = {7},
  year = {2011},
  month = {Oct},
  publisher = {American Physical Society},
  doi = {10.1103/PhysRevB.84.134502},
  url = {https://link.aps.org/doi/10.1103/PhysRevB.84.134502}
}

@article{https://doi.org/10.1002/chir.23361,
author = {Atzori, Matteo and Train, Cyrille and Hillard, Elizabeth A. and Avarvari, Narcis and Rikken, Geert L. J. A.},
title = {Magneto-chiral anisotropy: From fundamentals to perspectives},
journal = {Chirality},
volume = {33},
number = {12},
pages = {844-857},
doi = {https://doi.org/10.1002/chir.23361},
url = {https://onlinelibrary.wiley.com/doi/abs/10.1002/chir.23361},
eprint = {https://onlinelibrary.wiley.com/doi/pdf/10.1002/chir.23361},
year = {2021}
}

@article{Rikken1997,
author={Rikken, G. L. J. A.
and Raupach, E.},
title={Observation of magneto-chiral dichroism},
journal={Nature},
year={1997},
month={Dec},
day={01},
volume={390},
number={6659},
pages={493-494},
issn={1476-4687},
doi={10.1038/37323},
url={https://doi.org/10.1038/37323}
}

@article{Lin-2022,
   title={Zero-field superconducting diode effect in small-twist-angle trilayer graphene},
   volume={18},
   ISSN={1745-2481},
   url={http://dx.doi.org/10.1038/s41567-022-01700-1},
   DOI={10.1038/s41567-022-01700-1},
   number={10},
   journal={Nature Physics},
   publisher={Springer Science and Business Media LLC},
   author={Lin, Jiang-Xiazi and Siriviboon, Phum and Scammell, Harley D. and Liu, Song and Rhodes, Daniel and Watanabe, K. and Taniguchi, T. and Hone, James and Scheurer, Mathias S. and Li, J.I.A.},
   year={2022},
   month=aug, pages={1221–1227} }

@Article{Wu2022,
author={Wu, Heng
and Wang, Yaojia
and Xu, Yuanfeng
and Sivakumar, Pranava K.
and Pasco, Chris
and Filippozzi, Ulderico
and Parkin, Stuart S. P.
and Zeng, Yu-Jia
and McQueen, Tyrel
and Ali, Mazhar N.},
title={The field-free Josephson diode in a van der Waals heterostructure},
journal={Nature},
year={2022},
month={Apr},
day={01},
volume={604},
number={7907},
pages={653-656},
issn={1476-4687},
doi={10.1038/s41586-022-04504-8},
url={https://doi.org/10.1038/s41586-022-04504-8}
}

@article{10.1103/physrevb.62.648, 
year = {2000}, 
title = {{Control of the supercurrent in a mesoscopic four-terminal Josephson junction}}, 
author = {Sun, Qing-feng and Wang, Jian and Lin, Tsung-han}, 
journal = {Physical Review B}, 
issn = {1098-0121}, 
doi = {10.1103/physrevb.62.648}, 
pages = {648--660}, 
number = {1}, 
volume = {62}, 

}

@article{10.1103/physrevb.54.7366, 
year = {1996}, 
title = {{Hamiltonian approach to the transport properties of superconducting quantum point contacts}}, 
author = {Cuevas, J. C. and Martín-Rodero, A. and Yeyati, A. Levy}, 
journal = {Physical Review B}, 
issn = {1098-0121}, 
doi = {10.1103/physrevb.54.7366}, 
pmid = {9984360}, 
eprint = {cond-mat/9605023}, 
}

@article{10.1103/physrevb.52.8358, 
year = {1995}, 
title = {{dc Josephson current through a quantum dot coupled with superconducting leads}}, 
author = {Ishizaka, Satoshi and Sone, Jun’ichi and Ando, Tsuneya}, 
journal = {Physical Review B}, 
issn = {1098-0121}, 
doi = {10.1103/physrevb.52.8358}, 
pmid = {9979838}, 

number = {11}, 
volume = {52}, 

}

@article{PhysRevLett.131.027001,
  title = {Ubiquitous Superconducting Diode Effect in Superconductor Thin Films},
  author = {Hou, Yasen and Nichele, Fabrizio and Chi, Hang and Lodesani, Alessandro and Wu, Yingying and Ritter, Markus F. and Haxell, Daniel Z. and Davydova, Margarita and Ili\ifmmode \acute{c}\else \'{c}\fi{}, Stefan and Glezakou-Elbert, Ourania and Varambally, Amith and Bergeret, F. Sebastian and Kamra, Akashdeep and Fu, Liang and Lee, Patrick A. and Moodera, Jagadeesh S.},
  journal = {Phys. Rev. Lett.},
  volume = {131},
  issue = {2},
  pages = {027001},
  numpages = {6},
  year = {2023},
  month = {Jul},
  publisher = {American Physical Society},
  doi = {10.1103/PhysRevLett.131.027001},
  url = {https://link.aps.org/doi/10.1103/PhysRevLett.131.027001}
}

@article{
doi:10.1126/sciadv.abo0309,
author = {Margarita Davydova  and Saranesh Prembabu  and Liang Fu },
title = {Universal Josephson diode effect},
journal = {Science Advances},
volume = {8},
number = {23},
pages = {eabo0309},
year = {2022},
doi = {10.1126/sciadv.abo0309},
URL = {https://www.science.org/doi/abs/10.1126/sciadv.abo0309},
eprint = {https://www.science.org/doi/pdf/10.1126/sciadv.abo0309}
}

@article{MORIYA1982209,
title = {Evidence for the helical spin structure due to antisymmetric exchange interaction in Cr13NbS2},
journal = {Solid State Communications},
volume = {42},
number = {3},
pages = {209-212},
year = {1982},
issn = {0038-1098},
doi = {https://doi.org/10.1016/0038-1098(82)91006-7},
URL = {https://www.sciencedirect.com/science/article/pii/0038109882910067},
author = {T. Moriya and T. Miyadai}
}

@Article{Beach2021,
author={Beach, Alexander
and Reig-i-Plessis, Dalmau
and MacDougall, Gregory
and Mason, Nadya},
title={Asymmetric Fraunhofer spectra in a topological insulator-based Josephson junction},
journal={Journal of Physics: Condensed Matter},
year={2021},
month={Aug},
day={06},
publisher={IOP Publishing},
volume={33},
number={42},
pages={425601},
issn={0953-8984},
doi={10.1088/1361-648X/ac15d7},
url={https://doi.org/10.1088/1361-648X/ac15d7}
}

@Article{Falo2002,
author={Falo, F.
and Mart{\'i}nez, P. J.
and Mazo, J. J.
and Orlando, T. P.
and Segall, K.
and Tr{\'i}as, E.},
title={Fluxon ratchet potentials in superconducting circuits},
journal={Applied Physics A},
year={2002},
month={Aug},
day={01},
volume={75},
number={2},
pages={263-269},
issn={1432-0630},
doi={10.1007/s003390201325},
url={https://doi.org/10.1007/s003390201325}
}

@article{PhysRevE.61.2257,
  title = {Depinning of kinks in a Josephson-junction ratchet array},
  author = {Tr\'{\i}as, E. and Mazo, J. J. and Falo, F. and Orlando, T. P.},
  journal = {Phys. Rev. E},
  volume = {61},
  issue = {3},
  pages = {2257--2266},
  numpages = {0},
  year = {2000},
  month = {Mar},
  publisher = {American Physical Society},
  doi = {10.1103/PhysRevE.61.2257},
  url = {https://link.aps.org/doi/10.1103/PhysRevE.61.2257}
}

@ARTICLE{403209,
  author={Raissi, F. and Nordman, J.E.},
  journal={IEEE Transactions on Applied Superconductivity}, 
  title={Comparison of simulation and experiment for a Josephson fluxonic diode}, 
  year={1995},
  volume={5},
  number={2},
  pages={2943-2946},
  doi={10.1109/77.403209}
}

@article{PhysRevLett.87.077002,
  title = {Ratchet Effect: Demonstration of a Relativistic Fluxon Diode},
  author = {Carapella, G. and Costabile, G.},
  journal = {Phys. Rev. Lett.},
  volume = {87},
  issue = {7},
  pages = {077002},
  numpages = {4},
  year = {2001},
  month = {Jul},
  publisher = {American Physical Society},
  doi = {10.1103/PhysRevLett.87.077002},
  url = {https://link.aps.org/doi/10.1103/PhysRevLett.87.077002}
}

@ARTICLE{Gutfreund2023-ht,
  title    = "Direct observation of a superconducting vortex diode",
  author   = "Gutfreund, Alon and Matsuki, Hisakazu and Plastovets, Vadim and
              Noah, Avia and Gorzawski, Laura and Fridman, Nofar and Yang,
              Guang and Buzdin, Alexander and Millo, Oded and Robinson, Jason W
              A and Anahory, Yonathan",
  journal  = "Nat Commun",
  volume   =  14,
  number   =  1,
  pages    = "1630",
  month    =  mar,
  year     =  2023,
  address  = "England",
  language = "en"
}

@Article{Trahms2023,
author={Trahms, Martina
and Melischek, Larissa
and Steiner, Jacob F.
and Mahendru, Bharti
and Tamir, Idan
and Bogdanoff, Nils
and Peters, Olof
and Reecht, Ga{\"e}l
and Winkelmann, Clemens B.
and von Oppen, Felix
and Franke, Katharina J.},
title={Diode effect in Josephson junctions with a single magnetic atom},
journal={Nature},
year={2023},
month={Mar},
day={01},
volume={615},
number={7953},
pages={628-633},
issn={1476-4687},
doi={10.1038/s41586-023-05743-z},
url={https://doi.org/10.1038/s41586-023-05743-z}
}

@article{CAO2020100080,
title = {Overview and advances in a layered chiral helimagnet Cr1/3NbS2},
journal = {Materials Today Advances},
volume = {7},
pages = {100080},
year = {2020},
issn = {2590-0498},
doi = {https://doi.org/10.1016/j.mtadv.2020.100080},
url = {https://www.sciencedirect.com/science/article/pii/S2590049820300278},
author = {Y. Cao and Z. Huang and Y. Yin and H. Xie and B. Liu and W. Wang and C. Zhu and D. Mandrus and L. Wang and W. Huang}
}

@Article{Bauriedl2022,
author={Bauriedl, Lorenz
and B{\"a}uml, Christian
and Fuchs, Lorenz
and Baumgartner, Christian
and Paulik, Nicolas
and Bauer, Jonas M.
and Lin, Kai-Qiang
and Lupton, John M.
and Taniguchi, Takashi
and Watanabe, Kenji
and Strunk, Christoph
and Paradiso, Nicola},
title={Supercurrent diode effect and magnetochiral anisotropy in few-layer NbSe2},
journal={Nature Communications},
year={2022},
month={Jul},
day={23},
volume={13},
number={1},
pages={4266},
issn={2041-1723},
doi={10.1038/s41467-022-31954-5},
url={https://doi.org/10.1038/s41467-022-31954-5}
}

@article{PhysRevB.109.054508,
  title = {Chiral superconducting diode effect by Dzyaloshinsky-Moriya interaction},
  author = {Nunchot, Naratip and Yanase, Youichi},
  journal = {Phys. Rev. B},
  volume = {109},
  issue = {5},
  pages = {054508},
  numpages = {12},
  year = {2024},
  month = {Feb},
  publisher = {American Physical Society},
  doi = {10.1103/PhysRevB.109.054508},
  url = {https://link.aps.org/doi/10.1103/PhysRevB.109.054508}
}

@article{PhysRev.164.544,
  title = {Meissner Effect and Vortex Penetration in Josephson Junctions},
  author = {Goldman, A. M. and Kreisman, P. J.},
  journal = {Phys. Rev.},
  volume = {164},
  issue = {2},
  pages = {544--547},
  numpages = {0},
  year = {1967},
  month = {Dec},
  publisher = {American Physical Society},
  doi = {10.1103/PhysRev.164.544},
  url = {https://link.aps.org/doi/10.1103/PhysRev.164.544}
}

@article{nadeem2023superconductingdiodeeffect,
  title     = {The superconducting diode effect},
  author    = {Nadeem, Muhammad and Fuhrer, Michael S. and Wang, Xiaolin},
  journal   = {Nature Reviews Physics},
  volume    = {5},
  number    = {10},
  pages     = {558--577},
  year      = {2023},
  doi       = {10.1038/s42254-023-00632-w},
  publisher = {Nature Publishing Group},
  archivePrefix = {arXiv},
  eprint    = {2301.13564},
  primaryClass  = {cond-mat.supr-con},
  note      = {arXiv preprint at \url{https://arxiv.org/abs/2301.13564}},
}

@article{PhysRevB.103.245302,
  title = {Theory of the nonreciprocal Josephson effect},
  author = {Misaki, Kou and Nagaosa, Naoto},
  journal = {Phys. Rev. B},
  volume = {103},
  issue = {24},
  pages = {245302},
  numpages = {10},
  year = {2021},
  month = {Jun},
  publisher = {American Physical Society},
  doi = {10.1103/PhysRevB.103.245302},
  url = {https://link.aps.org/doi/10.1103/PhysRevB.103.245302}
}

@article{PhysRevX.12.041013,
  title = {General Theory of Josephson Diodes},
  author = {Zhang, Yi and Gu, Yuhao and Li, Pengfei and Hu, Jiangping and Jiang, Kun},
  journal = {Phys. Rev. X},
  volume = {12},
  issue = {4},
  pages = {041013},
  numpages = {11},
  year = {2022},
  month = {Nov},
  publisher = {American Physical Society},
  doi = {10.1103/PhysRevX.12.041013},
  url = {https://link.aps.org/doi/10.1103/PhysRevX.12.041013}
}

@article{
doi:10.1126/sciadv.ado1502,
author = {Fengshuo Liu  and Yuki M. Itahashi  and Shunta Aoki  and Yu Dong  and Ziqian Wang  and Naoki Ogawa  and Toshiya Ideue  and Yoshihiro Iwasa },
title = {Superconducting diode effect under time-reversal symmetry},
journal = {Science Advances},
volume = {10},
number = {31},
pages = {eado1502},
year = {2024},
doi = {10.1126/sciadv.ado1502},
URL = {https://www.science.org/doi/abs/10.1126/sciadv.ado1502},
eprint = {https://www.science.org/doi/pdf/10.1126/sciadv.ado1502}}

@article{Fukui_2018,
doi = {10.1088/1742-6596/1054/1/012027},
url = {https://dx.doi.org/10.1088/1742-6596/1054/1/012027},
year = {2018},
month = {jul},
publisher = {IOP Publishing},
volume = {1054},
number = {1},
pages = {012027},
author = {Fukui, Saoto and Kato, Masaru and Togawa, Yoshihiko and Sato, Osamu},
title = {Effects of chirality of a helical magnetic field on a superconductor},
journal = {Journal of Physics: Conference Series}
}

@article{PhysRevApplied.21.034011,
  title = {Magnetic-field-free nonreciprocal transport in graphene multiterminal Josephson junctions},
  author = {Zhang, Fan and Rashid, Asmaul Smitha and Tanhayi Ahari, Mostafa and de Coster, George J. and Taniguchi, Takashi and Watanabe, Kenji and Gilbert, Matthew J. and Samarth, Nitin and Kayyalha, Morteza},
  journal = {Phys. Rev. Appl.},
  volume = {21},
  issue = {3},
  pages = {034011},
  numpages = {9},
  year = {2024},
  month = {Mar},
  publisher = {American Physical Society},
  doi = {10.1103/PhysRevApplied.21.034011},
  url = {https://link.aps.org/doi/10.1103/PhysRevApplied.21.034011}
}

@article{PhysRevB.100.174511,
  title = {Two mechanisms of Josephson phase shift generation by an Abrikosov vortex},
  author = {Golod, T. and Pagliero, A. and Krasnov, V. M.},
  journal = {Phys. Rev. B},
  volume = {100},
  issue = {17},
  pages = {174511},
  numpages = {13},
  year = {2019},
  month = {Nov},
  publisher = {American Physical Society},
  doi = {10.1103/PhysRevB.100.174511},
  url = {https://link.aps.org/doi/10.1103/PhysRevB.100.174511}
}

@article{Rikken_Avarvari_2023, title={Comparing electrical magnetochiral anisotropy and chirality-induced spin selectivity}, volume={14}, DOI={10.1021/acs.jpclett.3c02546}, number={43}, journal={The Journal of Physical Chemistry Letters}, author={Rikken, G. L. and Avarvari, N.}, year={2023}, month={Oct}, pages={9727–9731}}

@Article{Qin2017,
author={Qin, F.
and Shi, W.
and Ideue, T.
and Yoshida, M.
and Zak, A.
and Tenne, R.
and Kikitsu, T.
and Inoue, D.
and Hashizume, D.
and Iwasa, Y.},
title={Superconductivity in a chiral nanotube},
journal={Nature Communications},
year={2017},
month={Feb},
day={16},
volume={8},
number={1},
pages={14465},
issn={2041-1723},
doi={10.1038/ncomms14465},
url={https://doi.org/10.1038/ncomms14465}
}

@Article{Legg2022,
author={Legg, Henry F.
and R{\"o}{\ss}ler, Matthias
and M{\"u}nning, Felix
and Fan, Dingxun
and Breunig, Oliver
and Bliesener, Andrea
and Lippertz, Gertjan
and Uday, Anjana
and Taskin, A. A.
and Loss, Daniel
and Klinovaja, Jelena
and Ando, Yoichi},
title={Giant magnetochiral anisotropy from quantum-confined surface states of topological insulator nanowires},
journal={Nature Nanotechnology},
year={2022},
month={Jul},
day={01},
volume={17},
number={7},
pages={696-700},
issn={1748-3395},
doi={10.1038/s41565-022-01124-1},
url={https://doi.org/10.1038/s41565-022-01124-1}
}

@article{Barron20021984,
author = {L.D. Barron and J. Vrbancich},
title = {Magneto-chiral birefringence and dichroism},
journal = {Molecular Physics},
volume = {51},
number = {3},
pages = {715--730},
year = {1984},
publisher = {Taylor \& Francis},
doi = {10.1080/00268978400100481},URL = {https://doi.org/10.1080/00268978400100481},eprint = { https://doi.org/10.1080/00268978400100481}}

@Article{Ideue2017,
author={Ideue, T.
and Hamamoto, K.
and Koshikawa, S.
and Ezawa, M.
and Shimizu, S.
and Kaneko, Y.
and Tokura, Y.
and Nagaosa, N.
and Iwasa, Y.},
title={Bulk rectification effect in a polar semiconductor},
journal={Nature Physics},
year={2017},
month={Jun},
day={01},
volume={13},
number={6},
pages={578-583},
issn={1745-2481},
doi={10.1038/nphys4056},
url={https://doi.org/10.1038/nphys4056}
}

@article{10.1063/5.0195229,
    author = {Birge, Norman O. and Satchell, Nathan},
    title = {Ferromagnetic materials for Josephson π junctions},
    journal = {APL Materials},
    volume = {12},
    number = {4},
    pages = {041105},
    year = {2024},
    month = {04},
    issn = {2166-532X},
    doi = {10.1063/5.0195229},
    url = {https://doi.org/10.1063/5.0195229},
    eprint = {https://pubs.aip.org/aip/apm/article-pdf/doi/10.1063/5.0195229/19866826/041105\_1\_5.0195229.pdf},
}

@Article{Clements2017,
author={Clements, Eleanor M.
and Das, Raja
and Li, Ling
and Lampen-Kelley, Paula J.
and Phan, Manh-Huong
and Keppens, Veerle
and Mandrus, David
and Srikanth, Hariharan},
title={Critical Behavior and Macroscopic Phase Diagram of the Monoaxial Chiral Helimagnet Cr1/3NbS2},
journal={Scientific Reports},
year={2017},
month={Jul},
day={26},
volume={7},
number={1},
pages={6545},
issn={2045-2322},
doi={10.1038/s41598-017-06728-5},
url={https://doi.org/10.1038/s41598-017-06728-5}
}

@Article{Sirica2020,
author={Sirica, N.
and Vilmercati, P.
and Bondino, F.
and Pis, I.
and Nappini, S.
and Mo, S.-K.
and Fedorov, A. V.
and Das, P. K.
and Vobornik, I.
and Fujii, J.
and Li, L.
and Sapkota, D.
and Parker, D. S.
and Mandrus, D. G.
and Mannella, N.},
title={The nature of ferromagnetism in the chiral helimagnet Cr1/3NbS2},
journal={Communications Physics},
year={2020},
month={Apr},
day={03},
volume={3},
number={1},
pages={65},
issn={2399-3650},
doi={10.1038/s42005-020-0333-3},
url={https://doi.org/10.1038/s42005-020-0333-3}
}

@Article{Ai2021,
author={Ai, Linfeng
and Zhang, Enze
and Yang, Jinshan
and Xie, Xiaoyi
and Yang, Yunkun
and Jia, Zehao
and Zhang, Yuda
and Liu, Shanshan
and Li, Zihan
and Leng, Pengliang
and Cao, Xiangyu
and Sun, Xingdan
and Zhang, Tongyao
and Kou, Xufeng
and Han, Zheng
and Xiu, Faxian
and Dong, Shaoming},
title={Van der Waals ferromagnetic Josephson junctions},
journal={Nature Communications},
year={2021},
month={Nov},
day={12},
volume={12},
number={1},
pages={6580},
issn={2041-1723},
doi={10.1038/s41467-021-26946-w},
url={https://doi.org/10.1038/s41467-021-26946-w}
}

@article{PhysRevB.95.035307,
  title = {Anomalous Fraunhofer interference in epitaxial superconductor-semiconductor Josephson junctions},
  author = {Suominen, H. J. and Danon, J. and Kjaergaard, M. and Flensberg, K. and Shabani, J. and Palmstr\o{}m, C. J. and Nichele, F. and Marcus, C. M.},
  journal = {Phys. Rev. B},
  volume = {95},
  issue = {3},
  pages = {035307},
  numpages = {11},
  year = {2017},
  month = {Jan},
  publisher = {American Physical Society},
  doi = {10.1103/PhysRevB.95.035307},
  url = {https://link.aps.org/doi/10.1103/PhysRevB.95.035307}
}

@Article{Mayer2020,
author={Mayer, William
and Dartiailh, Matthieu C.
and Yuan, Joseph
and Wickramasinghe, Kaushini S.
and Rossi, Enrico
and Shabani, Javad},
title={Gate controlled anomalous phase shift in Al/InAs Josephson junctions},
journal={Nature Communications},
year={2020},
month={Jan},
day={10},
volume={11},
number={1},
pages={212},
issn={2041-1723},
doi={10.1038/s41467-019-14094-1},
url={https://doi.org/10.1038/s41467-019-14094-1}
}

@Article{Assouline2019,
author={Assouline, Alexandre
and Feuillet-Palma, Cheryl
and Bergeal, Nicolas
and Zhang, Tianzhen
and Mottaghizadeh, Alireza
and Zimmers, Alexandre
and Lhuillier, Emmanuel
and Eddrie, Mahmoud
and Atkinson, Paola
and Aprili, Marco
and Aubin, Herv{\'e}},
title={Spin-Orbit induced phase-shift in Bi2Se3 Josephson junctions},
journal={Nature Communications},
year={2019},
month={Jan},
day={10},
volume={10},
number={1},
pages={126},
issn={2041-1723},
doi={10.1038/s41467-018-08022-y},
url={https://doi.org/10.1038/s41467-018-08022-y}
}

@Article{Pal2022,
author={Pal, Banabir
and Chakraborty, Anirban
and Sivakumar, Pranava K.
and Davydova, Margarita
and Gopi, Ajesh K.
and Pandeya, Avanindra K.
and Krieger, Jonas A.
and Zhang, Yang
and Date, Mihir
and Ju, Sailong
and Yuan, Noah
and Schr{\"o}ter, Niels B. M.
and Fu, Liang
and Parkin, Stuart S. P.},
title={Josephson diode effect from Cooper pair momentum in a topological semimetal},
journal={Nature Physics},
year={2022},
month={Oct},
day={01},
volume={18},
number={10},
pages={1228-1233},
issn={1745-2481},
doi={10.1038/s41567-022-01699-5},
url={https://doi.org/10.1038/s41567-022-01699-5}
}

@article{paterson2020tensile,
  title={Tensile deformations of the magnetic chiral soliton lattice probed by Lorentz transmission electron microscopy},
  author={Paterson, GW and Tereshchenko, AA and Nakayama, S and Kousaka, Y and Kishine, J and McVitie, Stephen and Ovchinnikov, AS and Proskurin, I and Togawa, Yoshihiko},
  journal={Physical Review B},
  volume={101},
  number={18},
  pages={184424},
  year={2020},
  publisher={APS}
}

@article{PhysRevResearch.6.L012046,
  title = {Signature of long-ranged spin triplets across a two-dimensional superconductor/helimagnet van der Waals interface},
  author = {Spuri, A. and Nikoli\ifmmode \acute{c}\else \'{c}\fi{}, D. and Chakraborty, S. and Klang, M. and Alpern, H. and Millo, O. and Steinberg, H. and Belzig, W. and Scheer, E. and Di Bernardo, A.},
  journal = {Phys. Rev. Res.},
  volume = {6},
  issue = {1},
  pages = {L012046},
  numpages = {6},
  year = {2024},
  month = {Mar},
  publisher = {American Physical Society},
  doi = {10.1103/PhysRevResearch.6.L012046},
  url = {https://link.aps.org/doi/10.1103/PhysRevResearch.6.L012046}
}

@article{PhysRevLett.132.216001,
  title = {Universal Spin Superconducting Diode Effect from Spin-Orbit Coupling},
  author = {Mao, Yue and Yan, Qing and Zhuang, Yu-Chen and Sun, Qing-Feng},
  journal = {Phys. Rev. Lett.},
  volume = {132},
  issue = {21},
  pages = {216001},
  numpages = {7},
  year = {2024},
  month = {May},
  publisher = {American Physical Society},
  doi = {10.1103/PhysRevLett.132.216001},
  url = {https://link.aps.org/doi/10.1103/PhysRevLett.132.216001}
}

@article{
doi:10.1126/science.1133239,
author = {Francoise Kidwingira  and J. D. Strand  and D. J. Van Harlingen  and Yoshiteru Maeno },
title = {Dynamical Superconducting Order Parameter Domains in Sr<sub>2</sub>RuO<sub>4</sub>},
journal = {Science},
volume = {314},
number = {5803},
pages = {1267-1271},
year = {2006},
doi = {10.1126/science.1133239},
URL = {https://www.science.org/doi/abs/10.1126/science.1133239},
eprint = {https://www.science.org/doi/pdf/10.1126/science.1133239}}

@article{Bouhon_2010,
doi = {10.1088/1367-2630/12/4/043031},
url = {https://dx.doi.org/10.1088/1367-2630/12/4/043031},
year = {2010},
month = {apr},
publisher = {},
volume = {12},
number = {4},
pages = {043031},
author = {Bouhon, Adrien and Sigrist, Manfred},
title = {Influence of the domain walls on the Josephson effect in Sr2RuO4},
journal = {New Journal of Physics}
}

@article{annurev:NoncentrosymmetricSuperconductors,
   author = "Yip, Sungkit",
   title = "Noncentrosymmetric Superconductors", 
   journal= "Annual Review of Condensed Matter Physics",
   year = "2014",
   volume = "5",
   number = "Volume 5, 2014",
   pages = "15-33",
   doi = "https://doi.org/10.1146/annurev-conmatphys-031113-133912",
   url = "https://www.annualreviews.org/content/journals/10.1146/annurev-conmatphys-031113-133912",
   publisher = "Annual Reviews",
   issn = "1947-5462",
   type = "Journal Article",
  }

@article{PhysRevLett.93.027003,
  title = {Evidence for a Novel State of Superconductivity in Noncentrosymmetric ${\mathrm{C}\mathrm{e}\mathrm{P}\mathrm{t}}_{3}\mathrm{S}\mathrm{i}$: A $^{195}\mathrm{P}\mathrm{t}$-NMR Study},
  author = {Yogi, M. and Kitaoka, Y. and Hashimoto, S. and Yasuda, T. and Settai, R. and Matsuda, T. D. and Haga, Y. and \ifmmode \bar{O}\else \={O}\fi{}nuki, Y. and Rogl, P. and Bauer, E.},
  journal = {Phys. Rev. Lett.},
  volume = {93},
  issue = {2},
  pages = {027003},
  numpages = {4},
  year = {2004},
  month = {Jul},
  publisher = {American Physical Society},
  doi = {10.1103/PhysRevLett.93.027003},
  url = {https://link.aps.org/doi/10.1103/PhysRevLett.93.027003}
}

@Article{Qiu2023,
author={Qiu, Gang
and Yang, Hung-Yu
and Hu, Lunhui
and Zhang, Huairuo
and Chen, Chih-Yen
and Lyu, Yanfeng
and Eckberg, Christopher
and Deng, Peng
and Krylyuk, Sergiy
and Davydov, Albert V.
and Zhang, Ruixing
and Wang, Kang L.},
title={Emergent ferromagnetism with superconductivity in Fe(Te,Se) van der Waals Josephson junctions},
journal={Nature Communications},
year={2023},
month={Oct},
day={23},
volume={14},
number={1},
pages={6691},
doi={10.1038/s41467-023-42447-4},
url={https://doi.org/10.1038/s41467-023-42447-4}
}

@article{PhysRevB.80.220502,
  title = {Direct dynamical coupling of spin modes and singlet Josephson supercurrent in ferromagnetic Josephson junctions},
  author = {Petkovi\ifmmode \acute{c}\else \'{c}\fi{}, I. and Aprili, M. and Barnes, S. E. and Beuneu, F. and Maekawa, S.},
  journal = {Phys. Rev. B},
  volume = {80},
  issue = {22},
  pages = {220502},
  numpages = {4},
  year = {2009},
  month = {Dec},
  publisher = {American Physical Society},
  doi = {10.1103/PhysRevB.80.220502},
  url = {https://link.aps.org/doi/10.1103/PhysRevB.80.220502}
}

@article{Chen2024HiddenStates,
  author  = {Chen, Shaowen and Park, Seunghyun and Vool, Uri and
             Maksimovic, Nikola and Broadway, David A. and Flaks, Mykhailo and
             Zhou, Tony X. and Maletinsky, Patrick and Stern, Ady and
             Halperin, Bertrand I. and Yacoby, Amir},
  title   = {Current induced hidden states in {J}osephson junctions},
  journal = {Nature Communications},
  year    = {2024},
  volume  = {15},
  number  = {1},
  pages   = {8059},
  doi     = {10.1038/s41467-024-52271-z}
}

@article{Yang2025FieldResilient,
  title   = {Field-resilient supercurrent diode in a multiferroic Josephson
             junction},
  author  = {Yang, Hung-Yu and Cuozzo, Joseph J. and Bokka, Anand Johnson and
             Qiu, Gang and Eckberg, Christopher and Lyu, Yanfeng and
             Huyan, Shuyuan and Chu, Ching-Wu and Watanabe, Kenji and
             Taniguchi, Takashi and Wang, Kang L.},
  journal = {Nature Communications},
  volume  = {16},
  pages   = {9287},
  year    = {2025},
  doi     = {10.1038/s41467-025-63698-3},
  url     = {https://www.nature.com/articles/s41467-025-63698-3},
}

@misc{KamraFu2024Conical,
  title         = {Nonreciprocal Josephson current through a conical magnet},
  author        = {{Johnsen Kamra}, Lina and Fu, Liang},
  year          = {2024},
  eprint        = {2409.00223},
  archivePrefix = {arXiv},
  primaryClass  = {cond-mat.supr-con},
  doi           = {10.48550/arXiv.2409.00223},
  url           = {https://arxiv.org/abs/2409.00223},
}

@article{Ghimire2013CrNbS2,
  title   = {Magnetic phase transition in single crystals of the chiral helimagnet
             {Cr}$_{1/3}${NbS}$_2$},
  author  = {Ghimire, N. J. and McGuire, M. A. and Parker, D. S. and Sipos, B. and
             Tang, S. and Yan, J.-Q. and Sales, B. C. and Mandrus, D.},
  journal = {Phys. Rev. B},
  volume  = {87},
  pages   = {104403},
  year    = {2013},
  doi     = {10.1103/PhysRevB.87.104403}
}

@article{Lee2024NbTiN,
  title   = {Penetration depth in dirty superconducting NbTiN thin films grown at
             room temperature},
  author  = {Lee, Yeonkyu and Yun, Jinyoung and Lee, Chanyoung and Sirena, M. and
             Kim, Jeehoon and Haberkorn, N.},
  journal = {Applied Physics A},
  volume  = {130},
  pages   = {504},
  year    = {2024},
  doi     = {10.1007/s00339-024-07650-0}
}

@article{Yu2005NbTiN,
  title   = {Fabrication of Niobium Titanium Nitride Thin Films With High
             Superconducting Transition Temperatures and Short Penetration
             Lengths},
  author  = {Yu, Lei and Singh, R. K. and Liu, Hongxue and Wu, Stephen Y. and
             Hu, Roger and Durand, D. and Bulman, John and Rowell, John M. and
             Newman, Nate},
  journal = {IEEE Transactions on Applied Superconductivity},
  volume  = {15},
  number  = {1},
  pages   = {44--47},
  year    = {2005},
  doi     = {10.1109/TASC.2005.844126}
}

@article{Tsuruta2016PRB,
  title   = {Phase diagram of the chiral magnet Cr$_{1/3}$NbS$_2$ in a magnetic
             field},
  author  = {Tsuruta, K. and Mito, M. and Deguchi, H. and Kishine, J. and
             Kousaka, Y. and Akimitsu, J. and Inoue, K.},
  journal = {Physical Review B},
  volume  = {93},
  pages   = {104402},
  year    = {2016},
  doi     = {10.1103/PhysRevB.93.104402}
}

@software{AlexanderBeach2025,
  author       = {Alexander Beach},
  title        = {Observation-of-a-Zero-Field-Josephson-
                   Diode-Effect-in-a-Helimagnet-Josephson-Junction:
                   Observation of a Zero-Field Josephson Diode Effect
                   in a Helimagnet Josephson Junction v1.0
                  },
  month        = sep,
  year         = 2025,
  publisher    = {Zenodo},
  version      = {v1.0},
  doi          = {10.5281/zenodo.17049491},
  url          = {https://doi.org/10.5281/zenodo.17049491},
}

@article{TokuraNagaosa2018,
  title   = {Nonreciprocal responses from non-centrosymmetric quantum materials},
  author  = {Tokura, Yoshinori and Nagaosa, Naoto},
  journal = {Nature Communications},
  volume  = {9},
  number  = {1},
  pages   = {3740},
  year    = {2018},
  doi     = {10.1038/s41467-018-05759-4},
  publisher = {Nature Publishing Group}
}
\clearpage
\appendix
\counterwithin*{figure}{part}
\counterwithin*{equation}{part}
\counterwithin*{table}{part}
\stepcounter{part}
\renewcommand{\thefigure}{A\arabic{figure}}
\renewcommand{\theequation}{A\arabic{equation}}
\renewcommand{\thetable}{A\arabic{table}}
\onecolumngrid
\section*{Theoretical Chiral Spin Structure Model and Vortex Simulations}\label{AppendixA}

\begin{figure}[h]
  \includegraphics[width=\columnwidth]{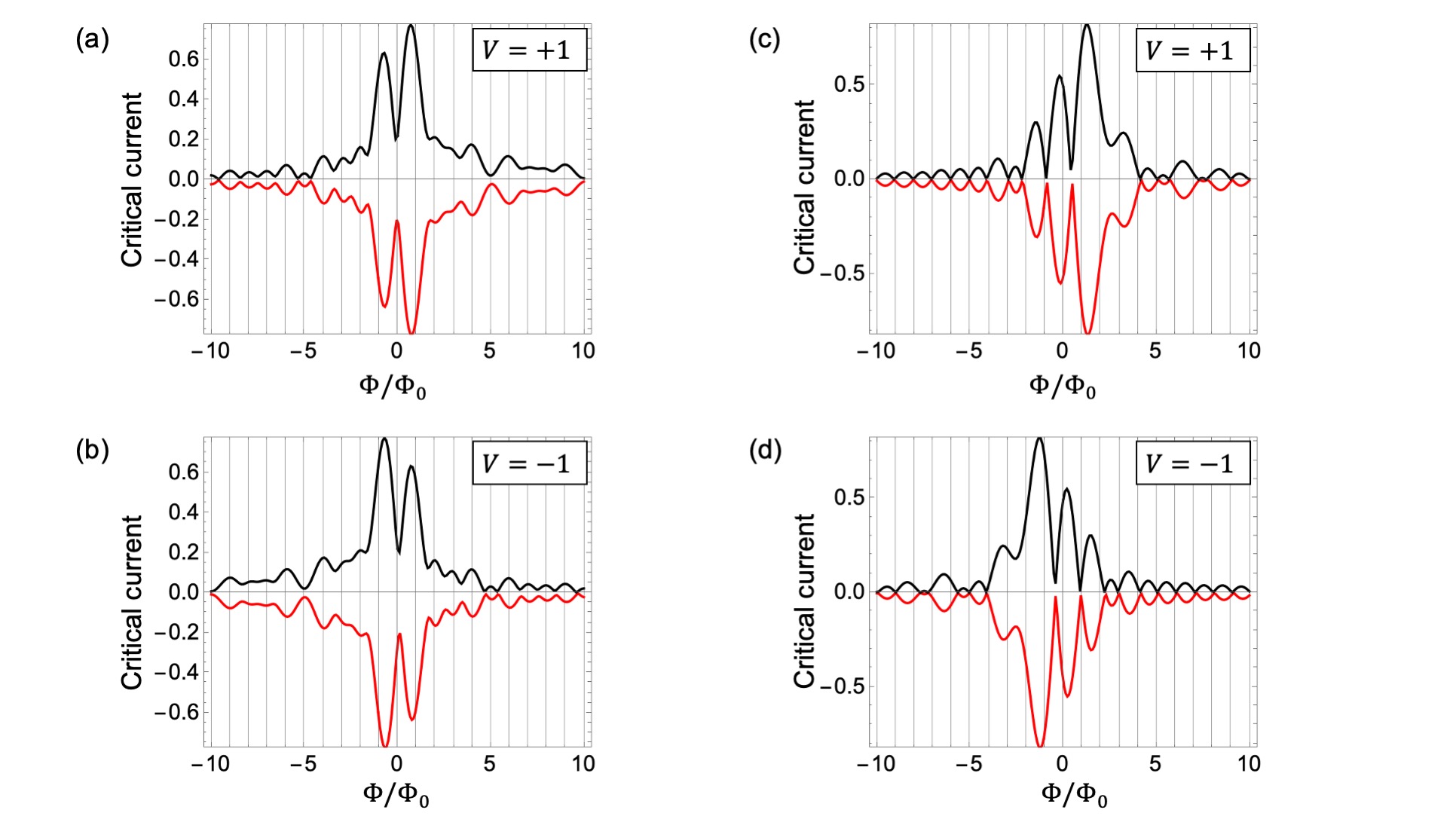}
   \caption{Fraunhofer patterns for a single vortex (a) antivortex (b) at
   $(x_v,y_v)/L_y=(0.01, 0.1)$, and double vortex (c) antivortex (d) at
   $(x_{v_1},y_{v_1})/L_y=(0.05, 0.1)$, $(x_{v_2},y_{v_2})/L_y=(0.05, -0.1)$ .}
   \label{image-A2}
\end{figure}

Fig.~\ref{image-A2} shows Fraunhofer patterns for a single vortex (a) antivortex (b) located at $(x_v,y_v)/L_y=(0.01, 0.1)$. The case of double vortex (c) antivortex (d) located at $(x_{v_1},y_{v_1})/L_y=(0.05, 0.1)$, $(x_{v_2},y_{v_2})/L_y=(0.05, -0.1)$ shows a greater shift of the maximum supercurrent compared to the single vortex case, which qualitatively match the experimental magnetic diffraction patterns.

We begin our modeling by constructing the Hamiltonian as follows:

\begin{equation}
H_{T} = H_{C} + H_{BCS}
\end{equation}
where $H_{C}$ represents the middle part of the junction with a square lattice structure and chiral spin structure, and $H_{BCS}$ denotes the left and right superconducting contacts of the junction. We project the sample onto the 2-dimensional $\hat{x},\hat{z}$ plane to account for the transport direction and helical axis. The dimensions are denoted as $N_{x} = 7$, $N_{z} = 11$. In real space, the first term of the Hamiltonian, which describes the central helimagnetic region of the junction, can be expressed as, 

\begin{equation}
    H_{C} =
    -t \sum_{i,j,\alpha}
    \left(c^{\dagger}_{i,j,\alpha} c_{i+1,j,\alpha} + \text{h.c.}\right)
    -t \sum_{i,j,\alpha}
    \left(c^{\dagger}_{i,j,\alpha} c_{i,j+1,\alpha} + \text{h.c.}\right)
    +\Delta_{c}\sum_{i,j}
    c^{\dagger}_{i,j}\left(\boldsymbol{\sigma}\cdot\hat{\mathbf{n}}_{j}\right)c_{i,j}.
\end{equation}

and the second term, which describes the s-wave pairing on the left and right contacts, as,

\begin{equation}
    H_{BCS} =
    \sum_{(i,j)\in L,R}
    \left[
    \Delta^{\ast}_{i,j}\,c^{\dagger}_{i,j,\uparrow}c^{\dagger}_{i,j,\downarrow}
    + \text{h.c.}
    \right].
\end{equation}

Here, $L$ and $R$ denote the sets of sites in the left and right superconducting contacts, $c_{i,j}=\left(c_{i,j\uparrow},c_{i,j\downarrow}\right)^T$ is the spinor annihilation operator at site $(i,j)$, and $c^{\dagger}_{i,j}$ is its Hermitian conjugate. The final term couples the conduction-electron spin to a local moment $\hat{\mathbf{n}}$, representing the magnetization, with coupling strength $\Delta_{c}$. To account for the helical spin structure, $\hat{\mathbf{n}}_{j} = \left(\sin(\theta)\cos(j\Delta\phi), \sin(\theta)\sin(j\Delta\phi), \cos(\theta)\right)$, where $\theta$ is the polar angle, $\Delta\phi$ is the inter-layer azimuthal rotation increment, and $j$ indexes layers along the $\hat{z}$-direction so that the azimuthal angle in layer $j$ is $j\Delta\phi$. $\Delta_{i,j}$ accounts for s-wave superconducting order parameter at $\left(i,j\right)$. To account for the semi-infinite superconducting contacts at the left ($x=0$) and right ($x=6$), we use the self-energy formalism in the Green’s function as follows in Eq.~\ref{eq:green}.\cite{10.1103/physrevb.62.648,10.1103/physrevb.54.7366,10.1103/physrevb.52.8358} In our model the exchange term proportional to $\Delta_c$ is present only in the central (Cr\sub{1/3}NbS\sub{2}) region, while the pairing potential $\Delta_{i,j}$ is nonzero only on the superconducting contact sites (and is set to zero in the central region). In practice, we integrate out the semi-infinite superconducting contacts, so their effect enters through the self-energies $\boldsymbol{\Sigma}_{L,R}$.

\begin{equation}
\label{eq:green}
G^{R}(\epsilon) = (\epsilon I + i\eta - H_{T})^{-1} = (\epsilon I + i\eta - H_{C} - \boldsymbol{\Sigma}_{L} - \boldsymbol{\Sigma}_{R})^{-1} 
\end{equation}

where 
\begin{equation}
\boldsymbol{\Sigma}_{L(R)}(\epsilon)= \frac{\tau_{L(R)}}{2}\,\beta(\epsilon)
\times\left(\begin{array}{cc}
I & i\frac{\Delta_{L(R)}}{\epsilon}\,\sigma_{y}\,e^{i \phi_{L(R)}}\\
-i\frac{\Delta_{L(R)}}{\epsilon}\,\sigma_{y}\,e^{-i \phi_{L(R)}} & I
\end{array}\right)
\end{equation}

In this context, $\epsilon$ is the energy, $I$ is the identity matrix of the appropriate dimension, $\eta$ is a small positive broadening parameter (not the diode efficiency from the main text),
$\tau_{L(R)}$ is the tunneling constant at the left (right) contact, $\beta(\epsilon)$ is the ratio between the superconducting and normal densities of states, $\sigma_{y}$ is a Pauli matrix, $\Delta_{L(R)}$ is the superconducting order parameter at the left (right) contact, and $\phi_{L(R)}$ is the superconducting phase at the left (right) contact.

In a Josephson junction, the superconducting order at the contact penetrates the middle part in order to lower the free energy. With the constructed Green's function, we utilize a self-consistent calculation to reach the ground state of the junction. The pairing superconducting potential in the central region is obtained from the energy spectrum of $H_{T}$ and then fed back into $H_{T}$ to calculate a new energy spectrum. 

After the self-consistent calculation converges, we obtain the equilibrium current--phase relation (CPR) by varying the superconducting phase difference $\varphi \equiv \phi_L-\phi_R$ between the two contacts. In equilibrium, the Josephson current can be expressed in terms of the Andreev bound-state spectrum $E_n(\varphi)$ as,

\begin{equation}
    I(\varphi) =
    \frac{2e}{\hbar}\sum_{n}\frac{\partial E_n(\varphi)}{\partial \varphi}
    \tanh\!\left(\frac{E_n(\varphi)}{2k_B T}\right),
    \label{eq:ABS_current}
\end{equation}

which reduces to the ground-state expression at $T\rightarrow 0$. From the CPR we extract the positive- and negative-bias critical currents as $I_{c+}=\max_{\varphi} I(\varphi)$ and $I_{c-}=\min_{\varphi} I(\varphi)$.

In Fig.~\ref{cpr} we show representative CPRs at $\theta = 60^{\circ}$ and $90^{\circ}$, corresponding to conical and coplanar helical spin textures,
respectively. For the conical parameters shown, we find $|I_{c-}|/I_{c+}\approx 0.89$, corresponding to $|\eta|\approx 0.058$ using $\eta=(I_{c+}-|I_{c-}|)/(I_{c+}+|I_{c-}|)$. The coplanar helical texture yields an essentially symmetric CPR (reciprocal response), while the conical texture yields an asymmetric CPR and a finite diode effect (nonzero $\eta$ in the definition used in the main text). The numerical parameters used in our calculations are summarized in Table~\ref{tab:numerical_parameters}.

In the main text we discuss two ingredients that can coexist in the same device: 1) a vortex-induced phase texture that shifts and distorts the
magnetic diffraction pattern, and 2) an intrinsic nonreciprocal current-phase relation (CPR) of the weak link. These can be treated in one expression by combining a vortex phase profile with an arbitrary local CPR $I_J(\vartheta)$.

We write the local phase difference along the junction as

\begin{equation}
\vartheta(y;\phi,\Phi)=\phi
+\frac{2\pi\Phi}{\Phi_0}\frac{y}{L_y}
+\phi_v(y),
\end{equation}

where $\phi$ is a global phase offset, $\Phi$ is the applied flux through the junction area, and $\phi_v(y)$ denotes the additional spatially varying phase contribution associated with trapped vortices (as modeled in the simulations above). The total current follows from,

\begin{equation}\label{app:combinedCPR}
I_{\mathrm{tot}}(\phi,\Phi)=\frac{1}{L_y}\int_{-L_y/2}^{L_y/2}
I_J\!\left(\vartheta(y;\phi,\Phi)\right)\,\mathrm{d}y.
\end{equation}
For each $\Phi$, the positive- and negative-bias critical currents are defined by extremizing over the global phase offset,

\begin{equation}
I_{c+}(\Phi)=\max_{\phi\in[0,2\pi)}\,I_{\mathrm{tot}}(\phi,\Phi),\qquad
I_{c-}(\Phi)=\min_{\phi\in[0,2\pi)}\,I_{\mathrm{tot}}(\phi,\Phi).
\end{equation}

The diode efficiency is then computed from the same definition used in the main text, $\eta(\Phi)=(I_{c+}-|I_{c-}|)/(I_{c+}+|I_{c-}|)$. As a concrete example of a CPR that breaks antisymmetry, one may consider

\begin{equation}
I_J(\vartheta)=I_1\sin\vartheta + I_2\sin(2\vartheta+\delta).
\label{eq:toyCPR}
\end{equation}

If $I_2\neq 0$ and $\delta\notin\{0,\pi\}$ (mod $2\pi$), then $I_J(\vartheta)\neq -I_J(-\vartheta)$, and a uniform junction ($\phi_v(y)=0$) can yield $I_{c+}\neq |I_{c-}|$ under the definitions above. In the presence of vortices, the same procedure applies with $\phi_v(y)\neq 0$. We include Eq.~(\ref{eq:toyCPR}) to make the combined calculation explicit, but we do not present additional combined numerical simulations here.

\begin{table}[!ht]
\centering
%\captionsetup{justification=centering}
\begin{tabular}{|c|c|c|}
\hline
\textbf{Parameter} & \textbf{Symbol} & \textbf{Value} \\
\hline
Number of lattice sites in the \( \hat{x} \) direction & \( N_{x} \) & 7 \\
\hline
Number of lattice sites in the \( \hat{z} \) direction & \( N_{z} \) & 11 \\
\hline
Hopping parameter & \( t \) & 1 \\
\hline
Exchange interaction strength & \( \Delta_{c} \) & 0.001 \\
\hline
Polar angle for spin orientation & \( \theta \) & 60° or 90° \\
\hline
Inter-layer azimuthal rotation increment for spin & \( \Delta\phi \) & 9° \\
\hline
Superconducting order parameter at the contacts & \( \Delta_{L(R)} \) & 0.1 \\
\hline
Transmission Coefficient of contacts& \( \tau_{L(R)} \) & 0.01 \\
\hline
Small positive value to avoid singularity & \( \eta \) & 0.0001 \\
\hline
Energy & \( \epsilon \) & -0.0024 \\
\hline
Superconducting phase at left contact & \( \phi_{L} \) & 0° \\
\hline
Superconducting phase at right contact & \( \phi_{R} \) & from 0° to 360° \\
\hline
\end{tabular}

\caption{Dimensionless numerical parameters used in the calculations (energies in units of $t$, lengths in units of $a$).}
\label{tab:numerical_parameters}
\end{table}

The numerical parameters used in our calculations are summarized in Table~\ref{tab:numerical_parameters}. In this Appendix we have used a standard tight-binding normalization in which energies are reported in units of the nearest-neighbor hopping $t$ and lengths are reported in units
of the lattice constant $a$ (i.e., we set $t=1$ and $a=1$ in the numerical code). These parameters are therefore not intended as a quantitative fit to the experimental devices; rather, they define a minimal model used to assess when a chiral-conical spin texture can generate a nonreciprocal CPR.

To connect the dimensionless parameters in Table~\ref{tab:numerical_parameters} to physical scales, let $\hbar=1$, and the lattice constant $a$ can be associated with an appropriate microscopic length (e.g., an atomic spacing or effective spacing), and the hopping $t$ with an electronic bandwidth scale. This way the superconducting gap entering the lead self-energies is $\Delta_{L(R)}\,t$ in energy units, and the coherence length is $\xi/a \sim v_F / (\pi \Delta_{L} a)$ in lattice units, where $v_F$ is the Fermi velocity of the effective band used in the model, which set by $t$ and the filling. Reported values for dirty NbTiN thin films are $\xi$ of a few nanometers and $\lambda$ of a few hundred nanometers, e.g. $\xi(0)\approx\SI{3.8}{\nano\meter}$ and $\lambda(0)\approx\SI{380}{\nano\meter}$, and $\lambda\approx$ \SI{200\pm20}{\nano\meter} at \SI{10}{\kelvin} with $\xi_{\mathrm{GL}}\approx$ \SI{3.8\pm0.3}{\nano\meter}.\cite{Yu2005NbTiN,Lee2024NbTiN} A fully quantitative mapping to the experimental NbTiN/Cr\sub{1/3}NbS\sub{2}/NbTiN junctions would require a material-specific band structure and device geometry, which is beyond the scope of the present minimal model. In this tight-binding model the parameters in Table~\ref{tab:numerical_parameters} are dimensionless once energies are measured in units of $t$. The spiral pitch is controlled by the inter-layer rotation angle $\Delta\phi$, and a full $2\pi$ rotation corresponds to $N_{\mathrm{pitch}}=2\pi/\Delta\phi$ lattice layers, and for $\Delta\phi=9^\circ$ and a \SI{48}{\nano\meter} helical period, $N_{\mathrm{pitch}}=40$ with an effective spacing along the helical axis of \SI{1.2}{\nano\meter}. The choice of $\Delta_c,\Delta_{L(R)},$ and $\tau_{L(R)}$ is intended to illustrate the qualitative dependence of the CPR on spin texture, i.e. the emergence of nonreciprocity when the magnetic moments have an out-of-plane component, rather than to provide a one-to-one quantitative fit to a specific device. A quantitative mapping to physical energy scales would require calibrating $t$ and $\Delta_c$ to the relevant band structure and exchange parameters of the proximitized Cr\sub{1/3}NbS\sub{2} weak link.

\end{document}